\documentclass[12pt]{article}
\usepackage[utf8]{inputenc}
\usepackage[T1]{fontenc}
\usepackage{amsmath,amssymb,tikz-cd}
\usepackage{upgreek}
\usepackage{nicematrix}
\usepackage{tikz}
\usetikzlibrary{arrows,chains,matrix,positioning,scopes,decorations,fadings}
\usepackage{simpler-wick}
\usepackage{hyperref}

\usepackage{comment}

\newcommand\mathd{\mathrm{d}}

\newcommand\BZ{\mathbb{Z}}

\newcommand\lam{\lambda}

\newcommand\bp{\mathbf{p}}
\newcommand\bx{\mathbf{x}}
\newcommand\by{\mathbf{y}}

\newcommand\sfZ{\mathsf{Z}}
\newcommand\sfF{\mathsf{F}}

\newcommand\sfA{\mathsf{A}}

\newcommand\sfC{\mathsf{C}}

\newcommand\CS{\mathrm{CS}}

\newcommand{\qL}{\rm qL}
\newcommand{\qG}{\rm qG}
\newcommand{\T}{\hat{T}}
\newcommand{\y}{\hat{y}}

\definecolor{bleudefrance}{rgb}{0.19, 0.55, 0.91}

\makeatletter
\tikzset{join/.code=\tikzset{after node path={%
\ifx\tikzchainprevious\pgfutil@empty\else(\tikzchainprevious)%
edge[every join]#1(\tikzchaincurrent)\fi}}}
\makeatother

\tikzset{>=stealth',every on chain/.append style={join},
         every join/.style={->}}
\tikzset{
    >=stealth',
    punkt/.style={
           rectangle,
           rounded corners,
           draw=black, very thick,
           text width=6.5em,
           minimum height=2em,
           text centered},
    pil/.style={
           ->,
           thick,
           shorten <=2pt,
           shorten >=2pt,}
}

\newcommand{\bea}{\begin{eqnarray}}
\newcommand{\eea}{\end{eqnarray}}
\newcommand{\be}{\begin{equation}}
\newcommand{\ee}{\end{equation}}

\DeclareMathAlphabet{\mathpzc}{OT1}{pzc}{m}{it}

\usepackage{amsthm}
\usepackage{amsmath}
\usepackage{amsfonts}
\usepackage{amssymb}

\theoremstyle{definition}

\newcommand{\ev}[1]{\Big\langle #1 \Big\rangle}

\newcommand{\qgPF}[1]{\sfZ^{\rm qG}_{U(1)}\left[#1 \right]}

\begin{document}

\begin{flushright} \small
 \end{flushright}
\smallskip
\begin{center} \Large
{\bf Lagrangian varieties from $q$-matrix models
}
 \\[12mm] \normalsize
{\bf Victor Mishnyakov${}^{a,b}$ and  Maxim Zabzine${}^{c,d}$} \\
[8mm]
{\small {\it $^a$ MIPT, Dolgoprudny 141701, Russia}}\\
{\small {\it $^b$ NRC ”Kurchatov Institute”, 123182, Moscow, Russia}}\\
{\small\it ${}^c$Department of Physics and Astronomy, Uppsala University,\\ Box 516, SE-75120 Uppsala, Sweden\\}
{\small\it ${}^d$Centre for Geometry and Physics, Uppsala University,\\
 Box 480, SE-75106 Uppsala, Sweden\\}
\end{center}
\vspace{5mm}
\begin{abstract} 
We consider three specific $q$-deformed matrix models (the Chern-Simons, $q$-Laguerre, and $q$-Gaussian matrix models)
which can be solved explicitly using the property of superintegrability. We show that one- and two-point functions of inverse characteristic polynomials
can be interpreted as quantizations of Lagrangian subvarieties in $(\mathbb{C}^*)^2$ and  $(\mathbb{C}^*)^4$, respectively. While such a geometric picture is expected for the Chern-Simons matrix model, our results for the $q$-Laguerre and $q$-Gaussian models are new and exhibit additional features. In particular, specific anti-symplectic birational involutions play an essential role in the construction. This reformulation provides a geometric framework for analyzing the semiclassical, large-$N$ expansion of matrix-model correlators.
This picture is suggestive of the geometry underlying open topological strings, although the two constructions are not identical and the precise relation between them remains to be understood. 
\noindent
\end{abstract}

\eject
\normalsize

\tableofcontents

\section{Introduction}

Matrix models provide a remarkably rich framework in mathematical physics. 
Despite their comparatively simple definition as finite-dimensional integrals, 
they capture highly non-trivial phenomena arising in gauge theory, string theory, enumerative geometry, statistical mechanics, and the theory of integrable systems.  
 One of the central approaches to matrix models is based on Ward identities or loop equations \cite{Mironov:1990im,dijkgraaf2002geometry,migdal1983loop,dijkgraaf1991loop}. 
 Instead of treating the matrix integral as the primary definition, one may characterize its partition/generating function as a solution of a system of differential or difference equations. This reformulation separates the algebraic structure of the model from analytic data such as the choice of integration contour and can turn otherwise difficult integral calculations into tractable algebraic problems. In favorable cases, Ward identities reduce to explicit recursion relations that determine all matrix-model correlation functions \cite{Lodin:2018lbz,Cassia:2021dpd,Mironov:2021yvg}. They therefore provide a natural starting point for studying models for which a direct analysis of the defining integral is difficult.

In many applications, matrix models arise in a $q$-deformed form, obtained by replacing ordinary algebraic and differential structures with their $q$-difference analogues. In some cases, the usual integral is also replaced by a discrete sum, known as a Jackson integral. Such models appear naturally in Chern-Simons theory, refined and topological string theory, supersymmetric gauge theories, knot theory, and quantum geometry. Their partition functions are typically constrained by $q$-Virasoro relations, or, more generally, by $q$-deformed $W$-algebras. This is the case, for example, for matrix models arising in the study of supersymmetric theories in three \cite{Beem:2012mb,Nedelin:2016gwu} and five dimensions \cite{Kimura_2018,aganagic2013gauge}. Despite the abundance of formal algebraic constructions based on free-field realizations, only a small number of models admit explicit solutions for all correlators in the form of well-defined recursion relations.

In this paper we study three explicitly solvable $q$-deformed matrix models: the Chern-Simons \cite{Tierz:2002jj}, $q$-Laguerre, and $q$-Gaussian models. Their exact solvability is a known consequence of superintegrability, which provides closed formulas for expectation values of Schur polynomials \cite{Mironov:2022fsr,Morozov:2018eiq,forrester2023q,Byun:2025qrv}. We use these established formulas as input rather than reconsidering superintegrability itself. Our aim is to reorganize the resulting Ward identities in a common geometric framework and to extend the analysis from the familiar rank-one quantum curves to higher-dimensional quantum varieties. The main new results are the explicit rank-two systems of $q$-difference equations, the identification of their reducible classical Lagrangian varieties, and the use of this structure to study the semiclassical large-$N$ regime.

At rank one, the normalized partition function of each of the models under consideration is uniquely determined by a single $q$-difference equation,
\begin{equation}
\mathsf{A}_{U(1)}[\y,\T]~\sfZ_{U(1)}[y]=0~.
\end{equation}
This property is special and should not be expected for a generic $q$-deformed matrix model. Taking the classical limit of the operator gives an algebraic equation
\begin{equation}
\mathcal{A}(y,T)=0~,
\end{equation}
whose vanishing locus defines a curve $\Sigma\subset(\mathbb{C}^*)^2$.  For the Chern--Simons model this picture is well-known and $\mathcal{A}(y,T)$ corresponds to the Hori-Vafa mirror curve for the resolved conifold \cite{Ooguri:1999bv,Hori:2000kt}. For the $q$-Laguerre and $q$-Gaussian models, by contrast, the appearance of similar geometric structures does not follow from a known topological-string construction.

The situation becomes considerably richer at higher rank, where the partition function depends on several variables. In the topological-string setting, this corresponds to placing a higher-rank theory, or equivalently a stack of branes, on a Lagrangian brane \cite{Marino:2004uf}. One then expects the full partition function to be annihilated by a system of $q$-difference operators whose classical symbols define a higher-dimensional Lagrangian variety \cite{Aganagic:2013jpa}. Explicit realizations of this expectation are presently available only in relatively simple cases, such as the conormal of the unknot and a small number of related examples. For general matrix models there is no independent geometric construction that predicts either the quantum operators or their underlying classical variety.

We therefore proceed in the opposite direction. Starting from the explicit matrix-model partition functions, we determine the $q$-difference operators that annihilate them and reconstruct the geometry from their classical limits. At rank two, for each of the three models, we find a system of three equations
\begin{equation}
\hat{A}_i[\y_1,\y_2,\T_1,\T_2]~
\sfZ_{U(2)}[y_1,y_2]=0~.
\end{equation}
The common zero locus of the corresponding classical symbols defines a two-dimensional Lagrangian variety
\begin{equation}
{\cal L}_{U(2)}\subset(\mathbb{C}^*)^4~.
\end{equation}
A new feature revealed by this analysis is that these varieties are generally reducible. In all our examples they have the schematic form
\begin{equation}
{\cal L}_{U(2)}
=
(\Sigma\times\Sigma)
\cup\Gamma_{\iota_1}
\cup\Gamma_{\iota_2}
\cup\cdots~.
\end{equation}
Thus there is always a component given by the direct product of two rank-one curves, but this component does not exhaust the full classical solution set. It is supplemented by components $\Gamma_{\iota_k}$ that can be identified with graphs of anti-symplectic involutions of the rank-one phase space $(\mathbb{C}^*)^2$. These additional components are invisible if one simply takes a product of the rank-one geometry.

The Chern-Simons model provides an important test of this construction. Its rank-$M$ partition function has a well-established topological-string interpretation as the partition function in the presence of $M$ branes wrapping the conormal of the unknot \cite{aganagic2004matrix}. It is therefore expected that the rank-one quantum curve reproduces the mirror curve of the resolved conifold. At rank two, the Lagrangian variety obtained directly from the matrix model agrees with the augmentation varieties studied in \cite{Aganagic:2013jpa,Ekholm:2024ceb}; in particular, the presence of several irreducible components is a known characteristic of such varieties. This agreement should be viewed as a non-trivial consistency check rather than as an entirely new geometric prediction. What is new is that the same structure is recovered directly and explicitly from the matrix-model Ward identities. This provides a useful laboratory for higher-rank knot and brane systems, for which the rank-one quantum $A$-polynomial \cite{Gukov:2003na} must be replaced by a structure governed by the skein algebra \cite{Ekholm:2019yqp}. These higher-rank techniques remain under active development, and explicit formulas are still scarce beyond examples such as the unknot and the Hopf link \cite{Ekholm:2024ceb,Ekholm:2024lir}.

The interpretation of the other two matrix models is less immediate and, for this reason, potentially more revealing. The $q$-Laguerre and $q$-Gaussian models arise from localization of three-dimensional $\mathcal{N}=2$ theories on $D_2\times S^1$, with one or two matter multiplets \cite{Cassia:2020uxy,beem2014holomorphic,Yoshida:2014ssa}. Difference equations are natural in this setting, but there is no a priori argument that their partition functions should be governed by mirror varieties of the type encountered in the knot and topological-string context. The new point is therefore not merely the existence of $q$-difference constraints, but the fact that their classical limits organize themselves into reducible Lagrangian varieties exhibiting the same general rank-two pattern as the Chern--Simons model.

Finally, the description of these matrix models as solutions associated with quantized Lagrangian varieties gives direct access to their semiclassical expansion. We consider the double-scaling limit
\begin{equation}
\hbar\longrightarrow0~,\qquad
N\longrightarrow\infty~,\qquad
a=q^N\ \text{fixed}~.
\end{equation}
This limit is substantially different from simply setting $q\to1$ at fixed $N$: the correlation functions and eigenvalue distributions retain a non-trivial dependence on the parameter $a=q^N$. The large-$N$ behavior at fixed $q$ remains poorly understood, despite several attempts to formulate it in terms of $q$-combinatorics \cite{Morozov:2020awm,wimberley2014towards,cohen2021moments}. The semiclassical double-scaling regime is more tractable and can be studied systematically using the classical Lagrangian variety and its quantization. Our analysis complements recent explicit results for moments and eigenvalue distributions in these models \cite{forrester2023q,byun2024q,Byun:2025hmz} and places them within a common quantum-geometric framework.

 The paper is organized as follows. In Section \ref{s:review}, we review some basic facts about $q$-matrix models and introduce the three matrix models studied throughout the paper. Section \ref{s:CS-model} is devoted to the Chern-Simons matrix model: we review the quantum and classical curves and present the explicit solution for $M=2$. In Section \ref{s:qL-model}, we consider the $q$-Laguerre matrix model, discussing its quantum and classical curves and deriving the explicit $M=2$ solution. Section \ref{s:qG-model} treats the $q$-Gaussian model, where we describe the quantum and classical curves for $M=1$ and obtain the explicit solution for $M=2$.
In Sections \ref{s:CS-model}--\ref{s:qG-model}, we also discuss specific symmetries of the classical curves that play a prominent role in the geometric interpretation of the $M=2$ solutions. Section \ref{s:higher-quantum-curve} contains the main result of the paper: we provide a geometric interpretation of the $M=2$ solutions as quantizations of Lagrangian subvarieties in  $(\mathbb{C}^*)^4$. In Section \ref{s:semiclassics}, we study the semiclassical expansion of our results and its relation to the underlying combinatorics. Finally, Section \ref{s:discussion} summarizes our results and discusses the limitations of the present approach, together with several open questions and directions for future work. We also include several appendices containing technical calculations and supplementary material supporting the discussions in the corresponding sections of the paper.

\section{$q$-deformed matrix models}\label{s:review}

In this section we define the models and fix conventions; we follow very closely \cite{Cassia:2025qga}. 
Matrix models define a measure on a collection of variables $\bx=(x_1,\ldots ,x_N)$, which are eigenvalues of a random matrix. Let $f(\bx)$ be a symmetric function of $\bx$ . The expectation value of this function in a matrix model with a potential $w(x)$ is defined as the following integral
\be\label{eq:evw}
 \ev{f}^{w} = \dfrac{1}{\sfZ^w}
 \int\limits_{\Gamma} \prod_{i=1}^{N} \mathd x_i  w(x_i)\, \Delta^2(\bx) \, f(\bx)~,
\ee
where $\Gamma$ is the integration contour and $\Delta(\bx)=\prod\limits_{i<j}(x_i-x_j)$ is the Vandermonde determinant. In general, the model is defined by the choice of the potential and integration contour. The normalization 
\begin{equation}
    \sfZ^w=\int\limits_{\Gamma} \prod_{i=1}^{N} \mathd x_i\,w(x_i)\Delta^2(\bx)
 \,
\end{equation}
is chosen in such a way that $\ev{1}^w=1$.

Instead of dealing with individual expectation values one works with generating functions. They are commonly referred to as matrix model partition functions and are defined as follows: 
\begin{equation}\label{eq:gen-func}
    \begin{split}
         \sfZ^w(\bp) :
 =&\, \frac{1}{\sfZ^w}\int_\Gamma \prod_{i=1}^N \mathd x_i\,w(x_i) 
 \exp\Big( \sum_{k\geq1}  \dfrac{p_k\, x_i^k}{k}  \Big) \Delta^2(\bx) \\
 =&\, \ev{ \exp\Big( \sum_{k\geq1}
  \frac{p_k}{k} \sum_{i=1}^N x_i^k \Big) }^w
 = \sum_\lam  \ev{ s_\lam(\bx) }^w s_\lam(\bp) \,.
    \end{split}
\end{equation}
The auxiliary parameters $p_k$ are called times.  They can either be thought of as independent variables or specialized to power sums of an auxiliary set of variables $\by$:
\begin{equation}
    p_k(\by)=\sum_{j=1}^M y_j^k\, .
\end{equation}

We call the corresponding partition function a rank $M$ partition function --  it depends on $M$ variables $\by$ and is denoted by:
\begin{equation}
     \sfZ^w_{U(M)}[\by] :=  \sfZ^w\Big( p_k(\by)=\sum_{j=1}^M y_j^k \Big) \,.
\end{equation}
In this case, the integrand can be rewritten by performing the sum in the exponent in \eqref{eq:gen-func}. As a result the rank $M$ partition function is nothing but an $M$-point function of inverse characteristic polynomials:
\begin{equation}
     \sfZ^w_{U(M)}[\by]= \ev{\prod_{j=1}^M\prod_{i=1}^N \dfrac{1}{1-x_i y_j}}^w
\end{equation}
  There are several peculiarities of $q$-deformed matrix models related to the choice of integration contour, which we briefly discuss below. In particular, some $q$-deformed matrix models are naturally formulated in terms of Jackson integrals rather than ordinary contour integrals.
The two definitions are related by modifying the potential by a special $q$-constant function $c_q(x)$ such that $c_q (qx) = c_q(x)$. Thus we can relate a contour integral to a Jackson integral as 
\begin{equation}
    \oint_\Gamma \prod_{i=1}^N \mathd x_i\,c_q(x_i)w(x_i) \Delta^2(\bx) f(\bx) = \int_{a}^b \prod_{i=1}^N \mathd_qx_i \,w(x_i)\Delta^2(\bx) f(\bx)~,
\end{equation}
where Jackson integrals are defined as the following sum
\be
 \int\limits_{0}^{a} \mathd_q x\,f(x)
 = (1-q)\sum_{n=0}^{\infty} (q^n a) f(q^n a) 
\ee
and
\be
 \int\limits_{a}^{b} \mathd_q x\,f(x)
 = \int\limits_{0}^{b} \mathd_q x\,f(x)-\int\limits_{0}^{a} \mathd_q x\,f(x)\,.
\ee
The function $c_{q}(x)$ is chosen so as to have poles at $x=q^{n}a$ and $x=q^{n}b$, with appropriately adjusted residues. This issue is discussed in detail in \cite{Lodin:2018lbz} and in the appendix of \cite{Cassia:2025qga}. We will not discuss it further here and will instead write the integral definition explicitly for each model.

One approach to solving matrix models is through Ward identities \cite{Mironov:1990im,Cassia:2021dpd}, which in our case are realized in the form of $q$-Virasoro constraints \cite{Lodin:2018lbz,Nedelin:2015mio,Aizawa:1990hx}. The partition function satisfies a set of $q$-difference equations in the $\bp$ variables. The equations can then be applied to solve for the partition function as a formal series in $\bp$. In this approach the problem of choosing the integration contour is recast in a different way. The dimension of the space of solutions depends on the potential and corresponds to the number of possible choices of contours with proper convergence properties \cite{Lodin:2018lbz,Cassia:2021dpd,Cassia:2020uxy}. 
In this paper, we restrict our attention to models for which the choice of contour is unique and the $q$-Virasoro constraints admit a unique solution. To construct this solution explicitly, we employ the recursion-operator approach of \cite{Cassia:2025qga}, in which suitable combinations of the $q$-Virasoro constraints are organized into recursion operators that determine the correlators recursively.

A crucial aspect of the models that we consider in this paper is that they are explicitly solvable. This property is also known as the superintegrability property \cite{Mironov:2022fsr,Cassia:2025qga}. In such models expectation values of Schur polynomials can be computed in closed form:
\begin{equation}
    \ev{s_{\lambda}(\bx)}^w=\sfC^w_{\lambda}(N) s_{\lambda}\left(p_k=\varphi_k^w \right)~.
\end{equation}
Here $\sfC^w_{\lambda}(N)$ is the so-called content product:
\begin{equation}
    \sfC^w_{\lambda}(N)=\prod_{(i,j) \in \lambda}g^w(q^{j-i})
\end{equation}
with the function $g^w$ depending on the model. The expression $s_{\lambda}\left(p_k=\varphi_k^w \right)$ is the image of the Schur function $s_{\lambda}$ under the specialization $ \varphi^{w}$ that fixes the power sums to a certain value:
\begin{equation}
    \varphi^{w}\left(p_k(\bf x)  \right) =\varphi^w_k~.
\end{equation}
Superintegrability means that the coefficients of the formal power series of the partition function are known explicitly. This is so due to the Cauchy identity, which allows one to express
    \begin{equation}\label{eq:gen-func2}
    \begin{split}
         \sfZ^w(\bp)=
 &\, \ev{ \exp\Big( \sum_{k\geq1}
  \frac{p_k}{k} \sum_{i=1}^N x_i^k \Big) }^w
 = \sum_\lam  \ev{ s_\lam(\bx) }^w s_\lam(\bp) \,.
    \end{split}
\end{equation}
Throughout the paper we use the following convention: when we write $s_{\lambda}(\bx)$ or $s_{\lambda}(\by)$ we understand the Schur function as a symmetric function of the collections of variables $\bx$ or $\by$, while when we write $s_{\lambda}(\bp)$ we understand it as a function of power sums. 

\subsection{Chern-Simons model}

The Chern-Simons or  Stieltjes–Wigert matrix model corresponds to the measure:
\begin{equation}
    w^{\CS}(x)=x^{\frac{\log r}{\log q}} \exp\left( -\dfrac{\log^2 x}{2 \log q} \right)~.
\end{equation}
The matrix model is defined as:
\begin{equation}
    \ev{f(\bx)}^{\CS}= \dfrac{1}{\sfZ^{\CS}}\int_{0}^{\infty} \prod_{i=1}^N \mathd x_i\, w^{\CS}(x_i) \Delta^2(\bx)  f(\bx)
\end{equation}
Here and in the models considered below, it is convenient to introduce the notation
\begin{equation}
    a=q^N \,.
\end{equation}
The $N$ dependence of $q$-deformed matrix models appears polynomially in $a$. For convenience we fix $r=q^{-\frac{1}{2}}a^{-1}$; then superintegrability takes the form:
\begin{equation}
    \ev{s_{\lambda}(\bx)}^{\rm CS} =  \left(\prod_{(i,j)\in \lambda}q^{j-i}(1-q^{j-i}a)  \right)\cdot s_{\lambda}\left(p_k=\dfrac{1}{1-q^k} \right)
\end{equation}
Thus we know the full generating function \eqref{eq:gen-func}. In particular we can write the rank one generating function explicitly. When $p_k=y^k$ 
\begin{equation}\label{eq:ZCSU1}
\sfZ^{\rm CS}_{U(1)}[y]= \ev{\dfrac{1}{\prod_{i=1}^N (1-x_i y) } }^{\CS}= \sum_{n=0}^{\infty}q^{\frac{n(n-1)}{2}} \dfrac{(a;q)_n}{(q;q)_n}  y^n
\end{equation}

\paragraph{Relation to Chern-Simons theory} The Chern-Simons matrix model was proposed first in  \cite{Marino:2002fk} 
 and then later derived via localization of the Chern-Simons theory in \cite{Kapustin:2009kz}. 
The expectation value of a Schur polynomial in the Chern-Simons model gives the HOMFLY-PT polynomial of the unknot \cite{Tierz:2002jj,Brini:2011wi}
\begin{equation}
    \ev{s_{\lambda}(\bx)}^{\rm CS} = H^{\rm unknot}_{\lambda}(a,q)\,.
\end{equation}
 Therefore the full rank $M$ partition function corresponds to the topological string partition function with a $U(M)$ theory living on the knot conormal. Therefore we have
\begin{equation}
    \sfZ^{\rm CS}_{U(1)}[y]=\sum_{n=0}^{\infty} H_{[n]}^{}(a,q)y^n \,.
\end{equation}
The rank two partition function can be rewritten in two ways. First, as a generating function of unknot HOMFLY-PT polynomials colored by two row representations:
    \begin{equation}
    \sfZ^{\rm CS}_{U(2)}[y_1,y_2]=\sum_{\lambda:l(\lambda)\leq 2} H^{\rm unknot}_{[\lambda_1,\lambda_2]}(a,q) s_{[\lambda_1,\lambda_2]}(y_1,y_2)
\end{equation}
On the other hand, in the Cauchy formula we could write:
\begin{equation}
    \exp\left( \sum_{k=1}^{\infty} \dfrac{(y_1^k+y_2^k)\left(\sum\limits_{i=1}^N x_i^k\right)}{k} \right)=\sum_{n_1,n_2 \geq 0} y_1^{n_1} y_2^{n_2} s_{[n_1]}(\bx)s_{[n_2]}(\bx)
\end{equation}
After taking the matrix model expectation value we obtain the Hopf link invariant:
\begin{equation}
    \ev{s_\lambda(\bx)s_{\mu}(\bx)}^{\rm CS}= H^{\text{Hopf}}_{\lambda,\mu}~,
\end{equation}
 which means that the rank two matrix model partition function is also a generating function of invariants of the Hopf link, which is colored by two symmetric representations $[n_1],[n_2]$:
\begin{equation}
    \sfZ^{\rm CS}_{U(2)}[y_1,y_2]=\sum_{n_1,n_2 \geq 0} H^{\text{Hopf}}_{[n_1],[n_2]} \, y_1^{n_1} y_2^{n_2}\,.
\end{equation}

\subsection{$q$-Laguerre and $q$-Gaussian models.}

 The $q$-Laguerre matrix model is defined by the weight
\begin{equation}\label{ql=potential}
    w^{\rm qL}(x)=x^{\frac{\log r}{\log q}} (qu_1x ;q)_{\infty}~.
\end{equation}
To define this model properly, one has to either deform the measure by a $q$-constant function or define it as a Jackson integral. In the latter approach we have:
\begin{equation}
    \ev{f(\bx)}^{\qL}= \dfrac{1}{\sfZ^{\qL}} \int\limits_{0}^{(u_1q)^{-1}} \prod_{i=1}^N \mathd_q x_i \,w^{\rm qL}(x_i) \Delta^2(\bx) f(\bx) 
\end{equation}
For convenience we choose $u_1=1$, in which case superintegrability is given by:
\begin{equation}
    \ev{s_{\lambda}(\bx)}^{\rm qL}= \prod_{(i,j) \in \lambda}(1-q^{j-i}a) \prod_{(i,j) \in \lambda}(1-r q^{j-i}a) \cdot s_{\lambda}\left(p_k= \dfrac{1}{1-q^k} \right)
\end{equation}
In rank one we have the following explicit form
\begin{equation}
    \sfZ_{U(1)}^{\rm qL}[y]=\sum_{n=0}^{\infty} \dfrac{(a;q)_{n}(ar;q)_n}{(q;q)_{n}} y^n~.
\end{equation}
In most of the discussion below we will also set $r=1$ for convenience; however, all of the results can be easily generalized to the case of generic $r$.
\\

The $q$-Gaussian model is defined by the weight:
\begin{equation}
    w^{\rm qG}(x)=(qx;q)_{\infty}(-qx;q)_{\infty}
\end{equation}
Similarly to the $\qL$ model all our results below could be extended to a generic Al-Salam-Carlitz weight 
\begin{equation}\label{weight-qG-general}
    w^{\rm ASC}(x)=(q u_1x;q)_{\infty}(qu_2x;q)_{\infty}
\end{equation}
with minimal modification. This matrix model arises as the result of localization on $D \times S^1$ 
of the  $\mathcal{N}=2$ $U(N)$ gauge theory coupled to two fundamentals with generic masses. 
For simplicity we mostly stick to the specific case of $u_1=-u_2=1$, which corresponds specifically to the $q$-deformation of the Gaussian matrix model. In this case the integral is given by:
\begin{equation}
    \ev{f(\bx)}^{\qG}= \dfrac{1}{\sfZ^{\qG}} \int\limits_{-q}^{q^{-1}} \prod_{i=1}^N \mathd_q x_i \,w^{\rm qG}(x_i) \Delta^2(\bx) f(\bx) 
\end{equation}
Superintegrability of the $q$-Gaussian matrix model gives:
\begin{equation}
    \ev{s_{\lambda}(\bx)}^{\rm qG} = \prod_{(i,j) \in \lambda}(1-aq^{j-i}) \cdot s_{\lambda}\left(p_k= \dfrac{1+(-1)^k}{1-q^k} \right)
\end{equation}
which at rank one has:
\begin{equation}
  \sfZ_{U(1)}^{\rm qG}[y] = \sum\limits_{n=0}^\infty \frac{(a;q)_{2n}}{(q^2; q^2)_{n}} y^{2n}
\end{equation}
Superintegrability for generic $u_1,u_2$ is given by 
\begin{equation}
    \ev{s_{\lambda}(\bx)}^{\rm ASC} = \prod_{(i,j) \in \lambda}(1-aq^{j-i}) \cdot s_{\lambda}\left(p_k= \dfrac{u_1^{-k}+u_2^{-k}}{1-q^k} \right)
\end{equation}

The $q$-Laguerre and $q$-Gaussian matrix models are deformations of their classical counterparts, the Gaussian and Laguerre ensembles defined as:
\begin{equation}
\begin{split}
        \ev{f(\bx)}^{\rm G} = \int_{-\infty}^{\infty} \prod_{i=1}^N dx_i \Delta^2(\bx) \exp\left( -\dfrac{1}{2}\sum_{i=1}^N x_i^2
 \right) f(\bx)~,
 \\
     \ev{f(\bx)}^{\rm L} = \int_{0}^{\infty} \prod_{i=1}^N dx_i \Delta^2(\bx)  \exp\left( -\sum_{i=1}^N x_i
 \right) f(\bx)~.
\end{split}
 \end{equation}
They can be recovered in the 
$q \rightarrow 1$ limit after appropriate rescaling. Namely, the potentials behave as
\begin{equation}\label{eq:scaling-weight}
    \begin{split}
     &\lim_{q \rightarrow 1}   w^{\qL}\left((1-q)x \right) =  e^{-x}~,
     \\
     &\lim_{q \rightarrow 1}   w^{\qG}\left((1-q)^{1/2}x \right) =  e^{-\frac{1}{2}x^2}~.
    \end{split}
\end{equation}

The same is true on the level of correlators. In contrast to the large $N$ semiclassical limit ($q \rightarrow 1$ and $a=q^N$ is fixed) which we mainly study in this paper, here one takes $N$ to be fixed and $a=q^N$. Hence when $q \rightarrow 1$ one also has $a \rightarrow1$. This is best seen in the superintegrability formulas. One has
\begin{equation}
\begin{split}\label{eq:SIlimit}
    &   \lim_{q \rightarrow 1} (1-q)^{|\lambda|\over 2} \ev{s_{\lambda}(\bx)}^{\qG} = \prod_{(i,j) \in \lambda}(N+j-i) s_{\lambda}\left( p_k = \delta_{k,2} \right)= \ev{s_{\lambda}(\bx)}^{\rm G}\\
    &   \lim_{q \rightarrow 1} (1-q)^{|\lambda|} \ev{s_{\lambda}(\bx)}^{\qL} = \prod_{(i,j) \in \lambda}(N+j-i)^2  s_{\lambda}\left( p_k = \delta_{k,1} \right)= \ev{s_{\lambda}(\bx)}^{\rm L}
\end{split}
\end{equation}
 Therefore, if we were to look at the partition function, then the limit would involve rescaling the $y$-variables in such a way as to cancel the factors in \eqref{eq:SIlimit}.

\subsection{Recursion operators.}

Partition functions of matrix models satisfy different kinds of equations. Among them are the recursion equations studied in \cite{Cassia:2025qga}. At least for those models that satisfy the superintegrability property, the partition function satisfies a special type of equation:
\begin{equation}
    \hat{\sfA}^{w}\cdot \sfZ^w(\bp)=0
\end{equation}
which we call a recursion equation, with $ \hat{\sfA}^{w}$ the recursion operator. These equations are certain combinations of $q$-Virasoro constraints \cite{Lodin:2018lbz,Nedelin:2015mio,Aizawa:1990hx}. Their main property is that they have a unique solution in the space of formal power series in the time variables. 
\\

In \cite{Cassia:2025qga}, the models considered have $t \neq q$ and  equations are expressed in terms of operators in the Fock representation of the quantum toroidal $\mathfrak{gl}_1$. In this paper we have $t=q$, which reduces the algebra to the $q$-$W_{1+\infty}$ algebra. Technically this means that the $  \hat{\sfA}^{w}$ are written in terms of modes of the vertex operators $x^\pm(z)=\sum\limits_{k\in\BZ}z^k x^\pm_{-k}$, with
\begin{equation}
    \label{eq:VertexOperatorsx}
 x^\pm(z) = \exp\left(\sum_{k\geq1}(1-q^{\mp k})z^k\frac{p_k}{k}\right)
 \exp\left(-\sum_{k\geq1}(1-q^{\pm k})z^{-k}\frac{\partial}{\partial p_k}\right)
\end{equation}
\\

These operators can also be rewritten in terms of the $y_i$ variables (see Appendix \ref{sec:AppendixDIMtoDAHA}). 

\section{Chern-Simons matrix model}\label{s:CS-model}

We start with the Chern-Simons model. There the recursion operator is given by \cite{Cassia:2025qga}
\begin{equation}
     \hat{\sfA}^{\CS} := (1-x^{-}_0)
 - \left((1-q^{-1})p_1-ax^{+}_{-1}\right) \,.
\end{equation}
The operator acts on the Fock space $\mathbb{C}[p_1,p_2,\ldots ]$, but can be consistently reduced to an operator acting on symmetric functions of a finite number of $\by$ variables using formulas in Appendix \ref{sec:AppendixDIMtoDAHA}. In particular at rank one we set $p_k=y^k$ and get
\begin{equation}\label{eq:AU1cs}
     \hat{\sfA}^{\CS}_{U(1)}=
-1 + \T^{-1} - \frac{\y}{q} + \frac{a}{q}\,\hat{y}\,\T~, 
\end{equation}
such that 
\begin{equation}
    \hat{\sfA}_{U(1)}^{\rm CS} \sfZ^{\rm CS}_{U(1)}[y]=0~,
\end{equation}
where by $\T$ and $\y$ we denote the $q$-Weyl algebra operators:
\begin{equation}
    \T \cdot f(y)=f(qy)\,, \quad  \y \cdot f(y)=yf(y)
\end{equation}
with the commutation relation:
\begin{equation}
    \T \y=q \y \T~.
\end{equation}
Another common normalization for the quantum curve is 
 \begin{equation}\label{eq-CS-curve-T2}
     \T  \hat{\sfA}^{\CS}_{U(1)} = 1 - \T -\y \left( 1  - a \T \right) \T
 \end{equation}
The obtained equation can be identified as the quantum mirror curve of the resolved conifold.  
The more familiar form appears if 
 we define new operators 
\begin{equation}
    \hat{X}_1= -  \y \T~,~~~~\hat{X}_2 = - \T~,~~~~~\hat{X}_2 \hat{X}_1 = q \hat{X}_1 \hat{X}_2
\end{equation}
 and the curve has the form
\begin{equation}
     \T  \hat{\sfA}^{\CS}_{U(1)} = 1 +\hat{X}_1 + \hat{X}_2 + a \hat{X}_1 \hat{X}_2~. 
 \end{equation}
   At this point we would like to stress that the CS matrix model with insertion of $\prod \frac{1}{1-x_i y}$ 
   prefers one particular framing \eqref{eq-CS-curve-T2}
    for the curve. To obtain the curve in a different framing we have to insert a different type of generating function into the matrix integral. 

Let us illustrate how the $q$-difference equation \eqref{eq:AU1cs} is solved by the series \eqref{eq:ZCSU1}. For this we rewrite the equation as follows
\begin{equation}
   \begin{gathered}
        \T\hat{\sfA}_{U(1)}^{\rm CS} \sfZ^{\rm CS}_{U(1)}[y]=0 
        \\
        \downarrow
        \\
       \sfZ^{\rm CS}_{U(1)}[y] = \dfrac{1}{ (1-\T)}\y(1-a \T)\T\sfZ^{\rm CS}_{U(1)}[y]
   \end{gathered}
\end{equation}
and use
\begin{equation}
  \dfrac{1}{ (1-\T)}\y(1-a \T)\T\cdot y^n = \dfrac{(1-aq^n)q^n}{1-q^{n+1}} y^{n+1}~.
\end{equation}
This is precisely the ratio between the successive coefficients of \eqref{eq:ZCSU1}.
\\

\subsection{Classical curve}\label{ss:CS-curve}

 The quantum curve (\ref{eq:AU1cs}) has been studied extensively in the literature starting from \cite{Hori:2000kt}
 and it 
  corresponds to the mirror curve of the resolved conifold written in a particular framing. Let us recall some basic 
   facts and set the conventions. 

 The quantum curve (\ref{eq:AU1cs}) can be understood as the quantization of the classical curve 
  defined by the following equation in $(\mathbb{C}^*)^2$
\begin{equation}\label{CS-classical-curve}
 \Sigma^{\rm CS}_a = \{ {\cal A}^{\rm CS} (y, T)
=
-1 + T^{-1} - y + ay\,T =0~,~~~(y,T) \in (\mathbb{C}^*)^2 \}~,     
\end{equation}
 where we use $T$ and $y$ (without hats) as the classical variables in $(\mathbb{C}^*)^2$ which is equipped 
  with the symplectic structure 
  \begin{equation}
      \omega = \frac{dy}{y} \wedge \frac{d T}{T}~. 
  \end{equation}
   Obviously $\Sigma^{\rm CS}_a$ is a Lagrangian variety in $((\mathbb{C}^*)^2, \omega)$. 
   $T$ can serve  as a good coordinate on  $\Sigma^{\rm CS}_a$ and we have the formula
   \begin{equation}
       y = \frac{1-T}{T(1-aT)}~,
   \end{equation}
 where for a given $T$ we have a unique $y$.
 Since $T \in \mathbb{C}^*$ and $y \in \mathbb{C}^*$ we have 
  \begin{equation}
      \Sigma^{\rm CS}_a \simeq \mathbb{C}^*_T \setminus  \Big \{ 1, \frac{1}{a} \Big \} \simeq \mathbb{P}^1 \setminus \Big \{ 0, \infty, 1, \frac{1}{a} \Big \}~. 
  \end{equation}
 Thus $\Sigma^{\rm CS}_a$ is $\mathbb{P}^1$ with four punctures.  

Let us discuss the underlying symplectic geometry further. Define the following map
 \begin{equation}\label{CS-def-involution}
        \iota~:~ (\mathbb{C}^*)^2~\rightarrow~(\mathbb{C}^*)^2~,~~~~~~ \iota (y, T)= (y', T') = \Big (y, \frac{1}{ay T} \Big )~,
   \end{equation}
   which is an anti-symplectomorphism of $(\mathbb{C}^*)^2$
   \begin{equation}
       \iota^* (\omega) = - \omega~. 
   \end{equation}
    This map is actually an involution: $\iota^2 = {\rm id}$ and it preserves ${\cal A}^{\rm CS} (y, T)$
\begin{equation}
    {\cal A}^{\rm CS} (y, T)= {\cal A}^{\rm CS} (y', T')~.
\end{equation}
  In particular if we restrict $\iota$ to ${\cal A}^{\rm CS} (y, T)=0$ then it maps $\Sigma^{\rm CS}_a$ to $\Sigma^{\rm CS}_a$. 
  The restriction of $\iota$ to the curve has the following interpretation.
 If we look at the projection onto the $y$ coordinate  
  \begin{equation}
      \Sigma^{\rm CS}_a~ \rightarrow ~\mathbb{C}_y^*~,
  \end{equation}
 then for every $y$ we have two solutions $T_\pm (y)$ which satisfy our equation. The exchange of two roots
  \begin{equation}\label{CS-echange-roots}
      T_+ (y)~\longleftrightarrow~T_-(y)
  \end{equation}
 corresponds to the involution $\iota$ acting on
 $\Sigma^{\rm CS}_a$. 
   
   We can define the quantum version of this involution \eqref{CS-def-involution} as follows
    \begin{equation}
        \iota_q (\y, \T) = (\iota_q(\y), \iota_q(\T)) = (\y, \frac{q^2}{a\y} \T^{-1})~,~~~~
    \end{equation}
   such that 
    \begin{equation}
        \iota_q (\hat{A}\hat{B}) = \iota_q (\hat{B}) \iota_q(\hat{A})~.
    \end{equation}
   Using that $\iota_q (1)=1$ we get
 \begin{equation}
    \iota_q (\T^{-1}) = \T \frac{a\y}{q^2} = a \frac{\y}{q} \T~. 
 \end{equation}
  It is straightforward to check that $\iota_q^2=1$ and we have
  \begin{equation}
 \iota_q\left(\hat{\sfA}^{\CS}_{U(1)}[y] \right)
=
-1 + \iota_q(T^{-1}) - \frac{y}{q} + \iota_q\left(\frac{a}{q}\,y\,T \right) = \hat{\sfA}^{\CS}_{U(1)}[y]   ~,    
\end{equation}
 which realises the obvious symmetry of the quantum curve 
 \begin{equation}
    \T^{-1}~\longleftrightarrow~\frac{a}{q} \y \T~.
 \end{equation}

\subsection{Rank $2$ operator}\label{ss:rank2-CS}

Going to rank $M=2$, after applying formulas from Appendix \ref{sec:AppendixDIMtoDAHA} we get
\begin{equation}  \label{eq:operCSrank2}
    \begin{split}
         \hat{\sfA}^{\rm CS}_{U(2)} =  & -q
   (q+1)-(\y_1+\y_2)+ \frac{q \left(q \y_1-\y_2\right)}{ \left(\y_1-\y_2\right)}\T_2^{-1}+\frac{q \left(\y_1-q \y_2\right)}{ \left(\y_1-\y_2\right)}\T_1^{-1} + \\
   & +\dfrac{a}{q}\left( \y_1\dfrac{q \y_1-\y_2}{\y_1-\y_2}  \T_{1} + \y_2\dfrac{\y_1-q \y_2}{\y_1-\y_2}  \T_{2} \right)~,
    \end{split}
\end{equation} 
where now $\T_1 \cdot f(\by)=f(qy_1,y_2)$ and $\T_2 \cdot f(\by)=f(y_1,qy_2)$.
 \begin{equation}\label{eq-CS-rank2}
     \hat{\sfA}^{\rm CS}_{U(2)} \sfZ^{\rm CS}_{U(2)}[y_1,y_2]=0~.
 \end{equation}

Let us discuss the solution of this equation. In \cite{Cassia:2025qga} a general procedure of solving this equation algebraically was presented. It follows that the solution to this equation that satisfies $\sfZ^{\rm CS}_{U(2)}[0,0]=1$ and is  symmetric under $y_1 \leftrightarrow y_2$ is unique. It turns out that symmetry is in fact a redundant constraint, as simply asking for the solution to be regular at zero automatically makes it symmetric. Instead of following the general logic here we present an explicit solution of the rank two equation in terms of the rank one partition functions. The emerging determinant structure is special to the $t=q$ case and is absent for generic $t$, as typically happens in $\beta$ ensembles.
\\

To solve the equation, notice that it can be rewritten as
 \begin{equation}\label{eq:CSconjugated}
 \begin{split}
   & (\y_1 - \y_2) \hat{\sfA}^{\rm CS}_{U(2)} \frac{1}{(\y_1-\y_2)}=\\
    & = q \Big ( - 1 - \frac{\y_1}{q} + q \T_1^{-1} + \frac{a}{q^2} \y_1 \T_1 \Big ) + q^2 \Big ( - 1 - \frac{\y_2}{q^2} + \T_2^{-1} + \frac{a}{q^2} \frac{\y_2}{q} \T_2 \Big ) \\
    & = q \Big ( - 1 - \frac{\y_2}{q} + q \T_2^{-1} + \frac{a}{q^2} \y_2 \T_2 \Big ) + q^2 \Big ( - 1 - \frac{\y_1}{q^2} + \T_1^{-1} + \frac{a}{q^2} \frac{\y_1}{q} \T_1 \Big ) ~.
           \end{split}
 \end{equation}
 The terms in parentheses look almost like the rank one quantum curve, with some of the variables $q$-shifted. This can be accounted for and we have  
 \begin{equation}
      \Big ( - 1 - \frac{\y_i}{q} + q \T_i^{-1} + \frac{a}{q^2} \y_i \T_i \Big )y_i    \sfZ^{\rm CS}_{U(1)}[ y_i, a]=0
  \end{equation}
   and 
   \begin{equation}
     \Big ( - 1 - \frac{\y_i}{q^2} + \T_i^{-1} + \frac{a}{q^2} \frac{\y_i}{q} \T_i \Big )    \sfZ^{\rm CS}_{U(1)}[ y_i q^{-1}, a q^{-1}] =0~.
   \end{equation}
   Thus the conjugated recursion operator in  \eqref{eq:CSconjugated} annihilates any linear combination of $ y_1 \sfZ^{\rm CS}_{U(1)}[y_1, a] \sfZ^{\rm CS}_{U(1)}[y_2 q^{-1}, a q^{-1}]$ and $  y_2 \sfZ^{\rm CS}_{U(1)}[y_2, a] \sfZ^{\rm CS}_{U(1)}[y_1 q^{-1}, a q^{-1}]$. To construct the kernel of the original $\hat{\sfA}^{\rm CS}_{U(2)}$ we would multiply this generic solution by $\frac{1}{(y_1-y_2)}$, hence obtaining
   \begin{equation}
       \hat{\sfA}^{\rm CS}_{U(2)} \left(\frac{c_1 \cdot  y_1 \sfZ^{\rm CS}_{U(1)}[y_1, a] \sfZ^{\rm CS}_{U(1)}[y_2 q^{-1}, a q^{-1}]+ c_2 \cdot y_2 \sfZ^{\rm CS}_{U(1)}[y_2, a] \sfZ^{\rm CS}_{U(1)}[y_1 q^{-1}, a q^{-1}]}{(y_1-y_2)} \right) =0~.
   \end{equation}
If we are interested in solutions that are relevant
to the matrix model, in the sense that they are power series in $y_1$ and $y_2$, then we enforce, additionally, that the solution be regular at $(y_1,y_2)=(0,0)$. This immediately 
fixes the relative coefficient $c_1=-c_2$. Hence, up to an overall constant, the solution that is regular at $(0,0)$ is unique and given by
   \begin{equation}\label{CS-M2-PF}
 \begin{split}
         \sfZ^{\rm CS}_{U(2)}[y_1, y_2] =& \frac{1}{y_1-y_2} \Big ( y_1 \sfZ^{\rm CS}_{U(1)}[y_1, a] \sfZ^{\rm CS}_{U(1)}[y_2 q^{-1}, a q^{-1}] \\& \phantom{\frac{1}{y_1-y_2} \Big (}-     y_2 \sfZ^{\rm CS}_{U(1)}[y_2, a] \sfZ^{\rm CS}_{U(1)}[y_1 q^{-1}, a q^{-1}] \Big )\\&= \frac{1}{y_1-y_2} \det \left ( \begin{array}{cc}
         y_1 \sfZ^{\rm CS}_{U(1)}[y_1,a]   & y_2 \sfZ^{\rm CS}_{U(1)}[y_2, a]  \\
       \sfZ^{\rm CS}_{U(1)}[y_1 q^{-1}, a q^{-1}]     &  \sfZ^{\rm CS}_{U(1)}[y_2 q^{-1}, aq^{-1}]
       \end{array} \right )~.
 \end{split}
   \end{equation}
   Thus this is the unique regular solution to the equation \eqref{eq-CS-rank2}. Our goal is to understand that $\sfZ^{\rm CS}_{U(2)}[y_1,y_2]$ appears through a quantization of some algebraic variety in $(\mathbb{C}^*)^4$ and the present discussion with the single operator $\hat{\sfA}^{\rm CS}_{U(2)}$ is not suitable for this quantization picture.
   In Section \ref{s:higher-quantum-curve} we will discuss how to trade the operator $\hat{\sfA}^{\rm CS}_{U(2)}$ for another set of operators which are suitable for the quantization picture. 

\section{$q$-Laguerre matrix model}\label{s:qL-model}

The recursion equation is
\be\label{eq:Nf1recusion}
 \hat{\sfA}^{\qL}\cdot \sfZ^{\qL}(\bp) = 0
\ee
where the recursion operator of the $q$-Laguerre matrix model reads \cite{Cassia:2025qga}
\begin{equation}
\label{eq:Nf1A1}
 \hat{\sfA}^{\qL} := (1-x^{-}_0)
 + u_1^{-1}\left( q^{-1} x^{-}_{-1}
 + a(1-q^{-1}) p_1
 + r a (1-q^{-1}) p_1
 - a^2 r x^{+}_{-1})\right)~.
\end{equation}

Using formulas from Appendix \ref{sec:AppendixDIMtoDAHA} at rank $M=1$ and setting $r=1,u_1=1$ we get the $q$-Laguerre model quantum curve, which has the following form
\begin{equation}\label{qL-qcurve-spec}
    \hat{\sfA}^{\rm qL}_{U(1)} = -a^2   \y\T- (q-2 a
   \y)+ (q-\y) \T^{-1}~,
\end{equation}
 such that 
\begin{equation}
    \hat{\sfA}_{U(1)}^{\rm qL} \sfZ^{\rm qL}_{U(1)}[y]=0~. 
\end{equation}
 To recover a more familiar form of the quantum curve we can introduce the operators
 \begin{equation}
     \hat{X}_1 = - \y~,~~~\hat{X}_2 = - \T~,~~~~~\hat{X}_2 \hat{X}_1 = q \hat{X}_1 \hat{X}_2
 \end{equation}
  such that the curve operator becomes
  \begin{equation}
      q^{-1} \T \hat{\sfA}^{\rm qL}_{U(1)} = 1 + \hat{X}_1 + \hat{X}_2 + 2 a \hat{X}_1 \hat{X}_2 + a^2 \hat{X}_1 \hat{X}_2^2~.
  \end{equation}
   In this expression we can recognize the standard form of the quantum mirror curve for a toric CY manifold (they can be generically toric orbifolds). However, the coefficients of the last two terms are not generic. 
  Let us see if we can make them generic by turning on other parameters in the matrix model.
If in the $q$-Laguerre matrix model \eqref{ql=potential} we restore the generic parameters $r$ and $u_1$
 then the quantum curve has the following form
\begin{equation}\label{qL-qcruve-general}
    \hat{\sfA}^{\rm qL}_{U(1)} = -a^2 r u_1^{-1}  \y\T- (q- a (r+1) u_1^{-1}
   \y)+ (q- u_1^{-1}\y) \T^{-1}~. 
\end{equation}
 If we introduce the operators
 \begin{equation}
     \hat{X}_1 = - u_1^{-1} \y~,~~~\hat{X}_2 = - \T~,~~~~~\hat{X}_2 \hat{X}_1 = q \hat{X}_1 \hat{X}_2\,,
 \end{equation}
  then we can rewrite the quantum curve as follows
  \begin{equation}\label{qL-general-qcurve}
      q^{-1} \T \hat{\sfA}^{\rm qL}_{U(1)} = 1 + \hat{X}_1 + \hat{X}_2 + a(r+1) \hat{X}_1 \hat{X}_2 + a^2 r \hat{X}_1 \hat{X}_2^2~.
  \end{equation}
 Now we see that the coefficients in front of two last terms are not fine tuned for generic values of $a$ and
  $r$. The parameter $u_1$ does not play any role in the curve. The quantum curve \eqref{qL-general-qcurve} 
   has the standard form of the Hori-Vafa mirror curve. Indeed, one can find a toric CY manifold which has 
    this curve as its mirror curve, see the next subsection. 

\subsection{Classical curve}
 
 For the $q$-Laguerre model the quantum curve \eqref{qL-qcurve-spec} can be understood as the quantization of the following classical curve in $(\mathbb{C}^*)^2$ with the standard symplectic structure
\begin{equation}
 \Sigma_a^{\rm qL} = \{ \mathcal{A}^{qL} (y,T) = -a^2 y T -(1-2ay) + (1-y)T^{-1}=0   ~,~~(y,T) \in (\mathbb{C}^*)^2 \} \,.
\end{equation}
  Here we assume that $a$ is generic and $a\neq 0,1$. 
 For the $ \Sigma_a^{\rm qL}$ curve $T$ can serve as a good coordinate 
 and we have
 \begin{equation}\label{qL-eqyT}
     y = \frac{1-T}{(1-aT)^2}
 \end{equation}
 and so it looks like $\mathbb{P}^1$ with 4 punctures. However it is actually a special case of $\mathbb{P}^1$ with 5 punctures 
  when two punctures collide. To see this we need to look at the classical curve with generic parameters. 
   The quantum curve \eqref{qL-qcruve-general} can be understood as the quantization of the following classical curve
\begin{equation}
 \Sigma_{a,r}^{\rm qL} = \{  -a^2 ry T -(1-(r+1)ay) + (1-y)T^{-1}=0    ~,~~(y,T) \in (\mathbb{C}^*)^2 \}~,
\end{equation}
where we set $u_1=1$ (since $u_1^{-1}$ is just an overall coefficient in front of $y$) and only $(r,a)$ are genuine 
 parameters in the problem. Now $T$ is a good coordinate on $\Sigma_{a,r}^{\rm qL}$ and we have  
  \begin{equation}
      y = \frac{1-T}{(1-a T)(1-a rT)}~.
  \end{equation}
  From this we can conclude that $\Sigma_{a,r}^{\rm qL}$ is $\mathbb{P}^1$ with 5 punctures
   \begin{equation}
     \Sigma^{\rm qL}_{a, r} \simeq \mathbb{P}^1 \setminus \Big \{ 0, \infty, 1, \frac{1}{a}, \frac{1}{a r} \Big \}~. 
 \end{equation}
  When $r\rightarrow 1$ two punctures collide and we get the curve $\Sigma_{a}^{\rm qL}$.
  In Appendix \ref{app:mirror-qL} we provide the mirror symmetry 
  interpretation of the quantum curve \eqref{qL-general-qcurve}.

 Let us restrict to the case $r=1$. 
  In Subsection \ref{ss:CS-curve} we have defined the anti-symplectic 
   involution \eqref{CS-def-involution} which preserves the curve $\mathcal{A}^{\rm CS}$. Now we are interested in a similar construction for the $q$-Laguerre curve $\Sigma^{\rm qL}_{a}$. Here
    the situation is trickier, we can define the map $\iota$ not on the 
 full $(\mathbb{C}^*)^2$ but on the open dense subset of $(\mathbb{C}^*)^2$
   \begin{align}\label{qL-involution-def}
    & \iota~:~ (\mathbb{C}^*\setminus \{1\}) \times \mathbb{C}^*~\rightarrow~(\mathbb{C}^*\setminus \{1\}) \times \mathbb{C}^*~, \nonumber \\
    &  \iota(y,T)= (y', T')= \Big  (y, \frac{y-1}{a^2 T y} \Big  )~,
\end{align}
  such that $\iota^2={\rm id}$ on this subset and it preserves $\mathcal{A}^{\rm qL}$
  \begin{equation}
      \mathcal{A}^{\rm qL} (y,T) = \mathcal{A}^{\rm qL}(y',T')~.
  \end{equation}
   We will refer to such $\iota$ as a birational anti-symplectic involution. We can restrict $\iota$ to the curve and it maps $\Sigma_a^{\rm qL}$ to itself. 
  The map from the curve to $y$
 \begin{equation}
     \Sigma_a^{\rm qL}~\rightarrow~\mathbb{C}^*_y
 \end{equation}
  is two-to-one  and the action of $\iota$ on $\Sigma_a^{\rm qL}$ can be interpreted as the exchange of two sheets under the above map.
  
  We can define the quantum version of the involution \eqref{qL-involution-def}
 \begin{equation}
    \iota_q (\y, \T) = \Big  ( \y, \T^{-1} \frac{q(\y-1)}{a^2 \y} \Big ) 
 \end{equation}
 with the property
 \begin{equation}
   \iota_q (\hat{A}\hat{B}) = \iota_q (\hat{B}) \iota_q (\hat{A}) ~. 
 \end{equation}
  We can derive the relation
  \begin{equation}
     \iota_q (\T^{-1}) = \frac{a^2\y}{q(\y-1)} \T 
  \end{equation}
   and $\iota_q^2 =1$. 
  Thus we have 
   \begin{equation}
       \iota_q \Big ( \hat{\sfA}^{\rm qL}_{U(1)} \Big ) =  \hat{\sfA}^{\rm qL}_{U(1)}~. 
   \end{equation}
   This is a formal quantum definition of $\iota_q$ and one needs to address the issue of how to deal with $\hat{y}-1$ and its inverse.
   We leave this question aside since we will not use the quantum version much in the forthcoming discussion.

\subsection{Rank $2$ operator}

 Now let us solve the $q$-Laguerre model for $\sfZ^{\rm qL}_{U(2)}[y_1,y_2]$. The logic is very similar to the Chern-Simons matrix model considerations in subsection 
 \ref{ss:rank2-CS}. 
At rank two  the $q$-Laguerre operator is given by the following expression
\begin{equation}
    \begin{split}
      \hat{\sfA}^{\rm qL}_{U(2)} =   -\frac{a^2 \y_1 \left(q \y_1-\y_2\right)}{q
   \left(\y_1-\y_2\right)} \T_1+\frac{a^2   \y_2
   \left(q \y_2-\y_1\right)}{q
   \left(\y_1-\y_2\right)}\T_2-\\-\left(-2 a \y_1-2
   a \y_2+q^2+q\right)+\\+\frac{
   \left(q-\y_2\right) \left(q \y_1-\y_2\right)}{
   \left(\y_1-\y_2\right)}\T_2^{-1}-\frac{
   \left(\y_1-q\right) \left(\y_1-q \y_2\right)}{
   \left(\y_1-\y_2\right)}\T_1^{-1}
    \end{split}
\end{equation}
 and we are interested in finding a solution 
\begin{equation}\label{eq-qL-rank2}
     \hat{\sfA}^{\rm qL}_{U(2)} \sfZ^{\rm qL}_{U(2)}[y_1,y_2]=0~,
 \end{equation}
 which is regular at $(0,0)$. 
The recursion equation can be solved in a similar way to the Chern-Simons model by observing the following
\begin{equation}
      \begin{split}
   &(\y_1-\y_2) \hat{\sfA}^{\rm qL}_{U(2)} \frac{1}{(\y_1 - \y_2)}  \Big ( (2a\y_1-q) - \frac{a^2}{q} \y_1 \T_1 + q(q-\y_1) \T_1^{-1} \Big ) \\&\quad + q \Big ( (2a q^{-1} \y_2 - q) - \frac{a^2}{q^2} \y_2 \T_2 + (q-\y_2 )\T_2^{-1} \Big )\\
   & = 1 \leftrightarrow 2
      \end{split}
  \end{equation}
 Let  $\sfZ^{\qL}_{U(1)}[y_i, a]$ solve  the rank one quantum curve equation for fixed $a$
   \begin{equation}
      \Big ( - a^2 \y_i \T_i + (2a\y_i - q) + (q-\y_i) \T_i^{-1} \Big ) \sfZ^{\qL}_{U(1)}[y_i, a]=0 \,.
   \end{equation}
 Then we have
    \begin{equation}
      \Big ( - \frac{a^2}{q} \y_i \T_i + (2a\y_i - q) +q (q-\y_i) \T_i^{-1} \Big )  y_i \sfZ^{\qL}_{U(1)}[y_i, a]=0 
      \end{equation}
 and 
    \begin{equation}
      \Big ( - \frac{a^2}{q^2} \y_i \T_i + (2\frac{a}{q}\y_i - q) +(q-\y_i) \T_i^{-1} \Big ) \sfZ^{\qL}_{U(1)}[y_i, a q^{-1}]=0 \,.
      \end{equation}
 Therefore, by similar reasoning, the regular solution can be written as
        \begin{equation}\label{qL-M2-PF}
 \begin{split}
        \sfZ^{\qL}_{U(2)}[y_1, y_2, a] &= \frac{1}{y_1-y_2} \Big ( y_1 \sfZ^{\qL}_{U(1)}[y_1, a] \sfZ^{\qL}_{U(1)}[y_2, a q^{-1}] \\&-  y_2 \sfZ^{\qL}_{U(1)}[y_2, a] \sfZ^{\qL}_{U(1)}[y_1, a q^{-1}] \Big )\\&= \frac{1}{y_1-y_2} \det \left ( \begin{array}{cc}
         y_1 \sfZ^{\qL}_{U(1)}[y_1,a]   & y_2 \sfZ^{\qL}_{U(1)}[y_2, a]  \\
       \sfZ^{\qL}_{U(1)}[y_1 , a q^{-1}]     &  \sfZ^{\qL}_{U(1)}[y_2, aq^{-1}]
       \end{array} \right )~.
 \end{split}
   \end{equation}
  This is the unique regular solution of the equation \eqref{eq-qL-rank2}.
   In Section \ref{s:higher-quantum-curve} we discuss how this can be obtained via quantization of a suitable variety in $(\mathbb{C}^*)^4$.

\section{$q$-Gaussian matrix model}\label{s:qG-model}

The recursion operator:
\be\label{eq:D2}
 \hat{\sfA}^{\qG} \cdot \sfZ^{\qG}(\bp)=0
\ee
for the $q$-Gaussian matrix model reads \cite{Cassia:2025qga}
\begin{multline}
 \hat{\sfA}^{\qG} := (1-x^{-}_0)
 + (u_1^{-1}+u_2^{-1}) \left(q^{-1} x^-_{-1}+a(1-q^{-1})p_1 \right) \\
 - (u_1 u_2)^{-1}\left(q^{-2} x^-_{-2}+a (1-q^{-2})p_2+a^2 x^{+}_{-2}\right)~.
\end{multline}

At rank $M=1$ and $u_{1}=-u_2=1$ we get the quantum curve of the $q$-Gaussian matrix model
 \begin{equation}\label{qG-curve-a}
     \hat{\sfA}^{\rm qG}_{U(1)} = ( a(q+1) \y^2 - q^2) + (q^2 - \y^2) \T^{-1} - a^2 q \y^2 \T~,
 \end{equation}
  such that 
  \begin{equation}
       \hat{\sfA}^{\rm qG}_{U(1)} \sfZ^{\rm qG}_{U(1)}[y]=0~. 
  \end{equation}
   If we look at the generic parameters $u_1$ and $u_2$ in \eqref{weight-qG-general} then 
 the quantum curve becomes
 \begin{align}\label{gq-qc-general}
      \hat{\sfA}^{\rm qG}_{U(1)} =& \Big ( - a(q+1) u_1^{-1} u_2^{-1} \y^2 + (u_1^{-1} + u_2^{-1}) a q \y - q^2 \Big )\nonumber \\&+ \Big ( q^2 -(u_1^{-1}+ u_2^{-1}) q \y + u_1^{-1} u_2^{-1}   \y^2 \Big ) \T^{-1} + u_1^{-1} u_2^{-1} a^2 q \y^2 \T~. 
 \end{align}
  This operator collapses to  \eqref{qG-curve-a} for the values $u_1=-u_2 = -1$. 
   Introducing the operators
   \begin{equation}
     \hat{X}_1 = - (u_1^{-1} + u_2^{-1}) \y~,~~~~\hat{X}_2 = - \T~,~~~\hat{X}_2 \hat{X}_1 = q \hat{X}_1 \hat{X}_2   
   \end{equation}
   we rewrite the quantum curve \eqref{gq-qc-general} as
   \begin{align}\label{qcurve-qG-general}
     q^{-2}\T   \hat{\sfA}^{\rm qG}_{U(1)} = 1 + \hat{X}_1 + \hat{X}_2 + a \hat{X}_1 \hat{X}_2
     + R \hat{X}_1^2 + (q+1) aR \hat{X}_1^2 \hat{X}_2 + q a^2 R \hat{X}_1^2 \hat{X}_2^2
   \end{align}
    where 
    \begin{equation}
        R = \frac{u_1 u_2}{(u_1 + u_2)^2}~.
    \end{equation}
  Thus the natural deformations of the matrix model do not give us the quantum curve with generic coefficients. The four coefficients are fine tuned to specific values and only $a$ and $R$ are
 arbitrary coefficients. At present we do not know how to introduce more parameters into the $q$-Gaussian matrix model and still be able to solve it explicitly. 
   
\subsection{Classical curve}

 The quantum curve \eqref{qG-curve-a} can be understood as the quantization of the following  classical curve
 \begin{equation}\label{def-qG-curve}
 \Sigma^{\rm qG}_a = \{ \mathcal{A}^{\rm qG}=  (2a y^2 -1) + (1-y^2) T^{-1} - a^2 y^2 T =0~,~~~(y,T) \in (\mathbb{C}^*)^2 \}~,
\end{equation}
 where we assume the standard symplectic structure on $(\mathbb{C}^*)^2$. The equation for the curve $\Sigma^{\rm qG}_a$ can be rewritten as
 \begin{equation}
    1-T=y^2(1-aT)^2\, 
\end{equation}
which gives
  \begin{equation}
      y^2=\frac{1-T}{(1-aT)^2}~.
  \end{equation}
This curve is a double cover of  the classical curve \eqref{qL-eqyT} from the previous section.  
 Actually, the curve $\Sigma^{\rm qG}_a$ can be identified, for generic $a\neq 0,1$, with a six-punctured
$\mathbb{P}^1$. To see this, introduce the variable
\begin{equation}
    z=y(1-aT).
\end{equation}
Using the defining equation of the curve, we obtain
\begin{equation}
    T=1-z^2,
\end{equation}
and consequently, on the curve we have
\begin{equation}
    y=\frac{z}{1-a+a z  ^2}.
\end{equation}
Thus the map
\begin{equation}
    z\longmapsto
    \left(
    T(z),y(z)
    \right)
    =
    \left(
    1-z^2,
    \frac{z}{1-a+a z^2}
    \right)
\end{equation}
defines an isomorphism between an appropriate punctured
$\mathbb{P}^1$ and $\Sigma^{\rm qG}_a$. Indeed, the inverse map is
simply
\begin{equation}
    z=y(1-aT),
\end{equation}
so different values of $z$ cannot correspond to the same point of
the curve. In particular, $z$ and $-z$ correspond to $(y,T)$ and
$(-y,T)$, respectively.
Since the curve is considered inside $(\mathbb{C}^*)^2$, we have to
remove the values of $z$ for which either $y$ or $T$ vanishes or
becomes singular. These are
\begin{equation}
    z=0,\qquad z=\infty,\qquad z=\pm1,\qquad
    z=\pm\sqrt{\frac{a-1}{a}}.
\end{equation}
Therefore,
\begin{equation}
    \Sigma^{\rm qG}_a
    \simeq
    \mathbb{P}^1\setminus
    \left\{
    0,\infty,\pm1,\pm\sqrt{\frac{a-1}{a}}
    \right\}.
\end{equation}
  Introducing the parameters $u_1$ and $u_2$ does not change much and 
   we still have $\mathbb{P}^1$ with six punctures with the location of two punctures depending on $u_1$ and $u_2$. 

The curve $\Sigma^{\rm qG}_a$ can be thought of as a degeneration of 
 the following curve 
     \begin{equation}\label{qG-ellipric}
      1- T = y^2 (1-a_1T) (1-a_2 T)~,
      \end{equation}
   which is a curve of genus $1$ (elliptic curve). If we define 
      \begin{equation}
          w= y (1-a_1 T) (1-a_2 T)~,
      \end{equation}
       then the curve becomes
       \begin{equation}
           w^2 = (1-T) (1-a_1 T) (1-a_2 T)~,
       \end{equation}
       and for the generic values of $a_i$ ($a_1 a_2 \neq 0, a_1 \neq a_2, a_1\neq 1, a_2 \neq 1$) the roots of the cubic polynomial on the RHS are distinct. Therefore
       we have an elliptic curve \eqref{qG-ellipric} with 6 punctures.
      The case of $a_1=a_2$ is its degeneration to $\mathbb{P}^1$. 
   At the moment we do not know how to realize the curve \eqref{qG-ellipric} in the matrix model.    

 Let us restrict our consideration to the curve $\Sigma^{\rm qG}_a$
  in \eqref{def-qG-curve}. We can define the birational involution 
   as follows
 \begin{align}\label{qG-involution-def}
    & \iota_\pm~:~ (\mathbb{C}^*\setminus \pm 1) \times \mathbb{C}^*~\rightarrow~(\mathbb{C}^*\setminus \pm 1) \times \mathbb{C}^*~, \nonumber \\
    &  \iota_\pm(y,T)=  \Big  (\pm y, \frac{y^2-1}{a^2 T y^2} \Big  )~,
\end{align}
 where we have to remove the divisor $y=\pm 1$ for the map to be well defined. These two maps have the following properties
 \begin{equation}
     \iota^*_\pm (\omega) = -\omega~,~~~~\iota^2_\pm = {\rm id}_{(\mathbb{C}^*\setminus \pm 1) \times \mathbb{C}^*}~.
 \end{equation}
  These involutions $\iota_\pm$ preserve $\mathcal{A}^{\rm qG}$
  \begin{equation}
      \mathcal{A}^{\rm qG} (y, T) = \mathcal{A}^{\rm qG}(y, T')~,~~~~
      \iota_\pm (y,T) = (\pm y,T')~.
  \end{equation}
 Thus these involutions descend to maps of the curve with two points removed
 \begin{equation}
     \iota_{\pm}~:~\left(\Sigma_a^{\rm qG} \setminus \{\pm 1, (2a-1)a^{-2}\}\right)~\rightarrow~\left(\Sigma_a^{\rm qG} \setminus \{\pm 1, (2a-1)a^{-2}\}\right)~.
 \end{equation}
  This is analogous to the symmetry that exchanges two roots
  \eqref{CS-echange-roots} for the CS model. 

   Let us describe another birational anti-symplectic map which preserves $\mathcal{A}^{\rm qG}$ and will play a role in the following discussion. Here we just present the map with its properties, for further explanation of this map we refer to  Appendix \ref{app:birational}. 
Let us define the constant $\alpha = \frac{a-1}{a}$ (remember that we assume $a$ to be generic, $a\neq 0$ and $a\neq 1$).  We define the map $\iota_s$ as 
   \begin{equation}\label{iotas-def}
       \iota_s (y, T) = (y', T')
   \end{equation}
 with the following explicit expressions
 \begin{align}
   &     T' = T \frac{\rho-\alpha^2}{\rho (1-\rho)} \label{eq-qG-invol-extra1}~,\\
   &  y'=\frac{\alpha y(1-aT)(1-\rho)}{
\rho(1-\rho)-aT(\rho-\alpha^2)} ~, \label{eq-qG-invol-extra2}
 \end{align}
 where we use the notation $\rho=y^2 (1-aT)^2$. This is a birational 
  map defined on an open dense subset $U \subset (\mathbb{C}^*)^2$, here we have to remove some divisors that correspond to zeros of the numerators and denominators in the transformations. 
  We can show the following properties
  \begin{equation}
      \iota_s^* \Big ( \frac{dT}{T} \wedge \frac{dy}{y} \Big ) =\frac{dT'}{T'} \wedge \frac{dy'}{y'}= -  \frac{dT}{T} \wedge \frac{dy}{y}
  \end{equation}
 and 
 \begin{equation}\label{ext-invol-prop1}
   \mathcal{A}^{\rm qG} (y,T) =  \mathcal{A}^{\rm qG} (y',T')   
 \end{equation}
  together with 
  \begin{equation}\label{ext-invol-prop2}
      y y' (1-aT) (1-aT') =\alpha~. 
  \end{equation}
 Moreover $\iota_s$ is an involution, $\iota_s^2={\rm id}_U$.

\subsection{Rank $2$ operator}

 Next we find the regular solution  $\sfZ^{\rm qG}_{U(2)}[y_1,y_2]$
  for the $q$-Gaussian matrix model. The logic below is similar to the Chern-Simons matrix model considerations in subsection 
 \ref{ss:rank2-CS}. 
At rank two  the $q$-Gaussian operator is  defined as
   \begin{align}
\hat{\sfA}_{U(2)}^{\rm qG}
&= -(1+q)\left(q^2 - a \y_1^2 - a \y_1 \y_2 + a q \y_1 \y_2 - a \y_2^2\right) \nonumber\\
&\quad - \frac{a^2 \y_1^2 (q \y_1 - \y_2)}{ (\y_1 - \y_2)}\, \T_1
+ \frac{a^2 \y_2^2 (-\y_1 + q \y_2)}{ (\y_1 - \y_2)}\, \T_2 \nonumber\\
&\quad - \frac{(-q + \y_1)(q + \y_1)(\y_1 - q \y_2)}{ (\y_1 - \y_2)}\, \T_1^{-1}
+ \frac{(q - \y_2)(q \y_1 - \y_2)(q + \y_2)}{ (\y_1 - \y_2)}\, \T_2^{-1} \nonumber\\
&\quad - a(1-q)(1+q)\, \y_1 \y_2~,
\end{align}
 where we get rid of an overall $q^{-2}$. We are interested in  a regular solution for the equation
 \begin{equation}\label{qG-rank2-eq}
       \hat{\sfA}_{U(2)}^{\rm qG} \sfZ^{\qG}_{U(2)}[y_1, y_2, a]=0~.
   \end{equation}
 
 We have the following relation
 \begin{equation}
     \begin{split}
   (\y_1 - \y_2) \hat{\sfA}_{U(2)}^{\rm qG} \frac{1}{(\y_1 - \y_2)} = 
    \Big ( a (q+1) \y_1^2 + a (q+1) \y_2^2 - (1+q)q^2 \Big ) \\   - a^2 \y_1^2 \T_1 - a^2 \y_2^2 \T_2 + q (q^2 -\y_1^2)\T_1^{-1} + q (q^2 - \y_2^2) \T_2^{-1} \\
    = \Big ( (a (q+1) \y_1^2 - q^2) + q (q^2-\y_1^2) \T_1^{-1} - a^2 \y_1^2 \T_1 \Big ) \\
    +
    q \Big ( (\frac{a}{q} (q+1) \y_2^2 - q^2) - \frac{a^2}{q^2} q \y_2^2 \T_2 + (q^2 - \y_2^2) \T_2^{-1} \Big ) \\
    = 1 \longleftrightarrow 2~.
     \end{split}
 \end{equation}
  Let us use the same tricks as before.
  If we define 
  \begin{equation}
      \Big ( ( a(q+1) \y_i^2 - q^2) + (q^2 - \y_i^2) \T_i^{-1} - a^2 q \y_i^2 \T_i\Big ) \sfZ^{\qG}_{U(1)}[y_i, a] =0 \,,
  \end{equation}
   then we have 
  \begin{equation}
      \Big ( ( a(q+1) y_i^2 - q^2) + q (q^2 - y_i^2) \T_i^{-1} - a^2  y_i^2 \T_i\Big ) y_i \sfZ^{\qG}_{U(1)}[y_i, a] =0 
  \end{equation}
   and 
    \begin{equation}
      \Big ( ( \frac{a}{q}(q+1) \y_i^2 - q^2) + (q^2 - \y_i^2) \T_i^{-1} - \frac{a^2}{q^2} q \y_i^2 \T_i\Big ) \sfZ^{\qG}_{U(1)}[y_i, a q^{-1}] =0~.
  \end{equation}
    Thus the only regular solution for \eqref{qG-rank2-eq} is given by
   \begin{equation}\label{def-qG-M2}
     \sfZ^{\qG}_{U(2)}[y_1, y_2, a] = \frac{1}{y_1-y_2} \Big ( y_1 \sfZ^{\qG}_{U(1)}[y_1,a] \sfZ^{\qG}_{U(1)}[y_2, aq^{-1}] -    y_2 \sfZ^{\qG}_{U(1)}[y_2,a] \sfZ^{\qG}_{U(1)}[y_1, aq^{-1}] \Big )~.
   \end{equation}
    In the next section we interpret it via the quantization of a suitable variety in $(\mathbb{C}^*)^4$. 
   
\section{Higher rank quantum varieties}\label{s:higher-quantum-curve}

  Before going into the detailed discussion of each matrix model let us outline the overall logic. 
 We denote the $U(M)$ partition function as follows
\begin{equation}
    \sfZ_{U(M)}[y_1,\ldots ,y_M]
\end{equation}
The rank $M$ partition function is annihilated by the rank $M$ recursion operator $\hat{\sfA}_{U(M)}$
\begin{equation}\label{quant-general}
    \hat{\sfA}_{U(M)}  \sfZ_{U(M)}[y_1,\ldots ,y_M]=0~,
\end{equation}
 where $\hat{\sfA}_{U(M)}$ is an operator which involves the operators $(\y_i, \T_i)$, $i=1,..., M$ such that $\T_i \y_j = q^{\delta_{ij}} \y_j \T_i$.  For rank one we have the following condition
\begin{equation}\label{quan-1curve}
    \hat{\sfA}_{U(1)} \sfZ_{U(1)}[y]=0~,
\end{equation}
 where the operator $\hat{\sfA}_{U(1)}$ involves $\y$ and $\T$. In the classical limit the operator collapses to the classical expression 
 \begin{equation}
     \lim\limits_{q\rightarrow 1} \hat{\sfA}_{U(1)}[\y,\T] = \mathcal{A}(y,T)~,
 \end{equation}
 where now the classical variables $(y, T) \in (\mathbb{C}^*)^2$. 
  If we define the symplectic structure on $(\mathbb{C}^*)^2$
  \begin{equation}
      \omega_{(\mathbb{C}^*)^2} = \frac{dT}{T} \wedge \frac{dy}{y}~,
  \end{equation}
 then we can interpret the quantum condition \eqref{quan-1curve} as the quantization of the following classical curve
 \begin{equation}
     \Sigma = \{ \mathcal{A}(y,T) =0\} \subset (\mathbb{C}^*)^2~,
 \end{equation}
 which is obviously a 1-dimensional Lagrangian submanifold in $(\mathbb{C}^*)^2$. 
 Here we denote the classical counterparts of the $\y$ and $\T$ operators by $(y,T)$, respectively. We have discussed the classical curves and their quantization for three different matrix models 
  in previous sections.

 The rank one case $M=1$ is simpler since the recursion 
  operator $\hat{\sfA}_{U(1)}$ can be interpreted right away as quantization of the Lagrangian variety $\mathcal{A}(y,T)=0$ in
   $(\mathbb{C}^*)^2$. For rank $M \geq 2$, this simple picture fails, and the single condition \eqref{quant-general}
   does not correspond  to the quantization of a Lagrangian variety in $(\mathbb{C}^*)^{2M}$. Ideally we would like to replace \eqref{quant-general} by the set of quantum conditions 
   \begin{equation}
    \hat{A}_i~ \sfZ_{U(M)}[y_1,\ldots ,y_M]=0~,
   \end{equation}
    such that these conditions can be interpreted as quantization of 
     Lagrangian variety (possibly with many components) in $(\mathbb{C}^*)^{2M}$. Unfortunately we do not know how to solve 
      this problem for the general rank $M$. In what follows we provide the detailed solution of this problem for $M=2$. 

   Let us outline the result for $M=2$. For all three matrix models 
    we will find 3 operators $\hat{A}_{i}$, $i=1,2,3$ which annihilate 
     $\sfZ_{U(2)}[y_1, y_2]$
     \begin{equation}
    \hat{A}_i~ \sfZ_{U(2)}[y_1,y_2]=0~,~~~i=1,2,3
   \end{equation} 
   and moreover $\sfZ_{U(2)}[y_1,y_2]$ will be the only regular solution of these 3 equations. These quantum conditions can be interpreted as quantization of a Lagrangian variety in $(\mathbb{C}^*)^4$ which has many components and is organized as 
\begin{equation}
    {\cal L}_{U(2)} = (\Sigma \times \Sigma) \cup \Gamma_{\iota_1} \cup  ...~,
\end{equation}   
 where the first component corresponds to $\{ \mathcal{A}(y_1, T_1)=0, \mathcal{A}(y_2, T_2)=0 \}$. The additional components $\Gamma_{\iota_i}$ correspond to the graphs of anti-symplectic involutions which we are going to describe. The concrete structure of 
  ${\cal L}_{U(2)}$ is model dependent. At the operator level 
   the skein recursion operator can be expressed through the operators 
    $\hat{A}_i$ as
   \begin{equation}
        \hat{\sfA}_{U(2)} = \sum_{i=1}^3  \hat{r}_i \hat{A}_i~,
    \end{equation}
 where $\hat{r}_i$ are operators by themselves. We find these explicit 
  operatorial relations for CS and qL models, but not for the qG model.

\subsection{Chern-Simons model}

We start with the Chern-Simons model. Explicit rank 2 difference equations that annihilate the topological string partition function can be found, for example, in refs \cite{Aganagic:2013jpa,Ekholm:2024ceb}, written with a specific choice of framing. The Chern-Simons matrix model lands in a different framing; hence, our version of these equations looks slightly different, but structurally they are the same.
Namely, we find that the partition function \eqref{CS-M2-PF} is annihilated by three operators $\hat{A}^{\rm CS}_i \, ,\ \ i=1,2,3$:
    \begin{equation}\label{eq:AiCS}
        \hat{A}_i^{\CS} \sfZ_{U(2)}^{\rm CS}[y_1,y_2]=0
    \end{equation}
 where 
 \begin{align}\label{eq:QVCS}
   &     \hat{A}^{\rm CS}_1 \;=\; \hat{\sfA}_{U(1)}^{\rm CS}[\y_1, \T_1]
 \;-\; \hat{\sfA}_{U(1)}^{\rm CS}[\y_2, \T_2]~,\nonumber \\
  & 
  \hat{A}^{\rm CS}_2\;=\;\hat{\sfA}^{\CS}_{U(1)}[\y_1,\T_1] (aq^{-1} \y_2 \T_1 \T_2 -1)~,
  \\
&
\hat{A}^{\rm CS}_3 \;=\;\hat{\sfA}^{\CS}_{U(1)}[\y_2,\T_2] (aq^{-1} \y_1 \T_1 \T_2 -1)\nonumber
\end{align}
and $\hat{\sfA}_{U(1)}^{\rm CS}[\y_i,\T_i]$ denotes the rank-one curve \eqref{eq:AU1cs}, with the operators entering its definition specified in the brackets.
   Moreover, as we demonstrate below, this set of operators is minimal in some sense. Specifically, given certain boundary conditions, it allows us to determine the matrix model partition function uniquely. 
The equations \eqref{eq:AiCS} are invariant under the exchange of $(\y_1, \T_1) \leftrightarrow (\y_2, \T_2)$.

First, let us prove that the rank 2 partition function is indeed annihilated by these operators. For the $  \hat{A}^{\rm CS}_1 $ operator we utilize the determinant representation \eqref{CS-M2-PF}. After applying the operator we are left with the complicated expression in terms of $\sfZ^{\CS}_{U(1)}[y_i]$ and $\sfZ^{\CS}_{U(1)}[y_i,a q^{-1}]$. Parts of it can be simplified by taking into account that the rank one partition functions are annihilated by the quantum curve. The rest can be reduced by using the contiguous relation
\begin{equation}
    \sfZ^{\CS}_{U(1)}[q^{-1 }y_i,a q^{-1}]- \sfZ^{\CS}_{U(1)}[y_i,a q^{-1}]= \dfrac{q-a}{q^2} y_i\sfZ^{\CS}_{U(1)}[y_i] ~.
\end{equation} 
We prove this relation and sketch the transformations in Appendix \ref{sec:AppendixSplittingProof}.
To prove that the second and third operators annihilate the partition function we should work out the properties of the following operators
\begin{equation}\label{eq:CSoperO}
  \begin{split}
\hat{\mathcal{O}}^{\rm CS}_{1}=(\y_1 aq^{-1} \T_2 \T_1-1)~,
        \\
    \hat{\mathcal{O}}^{\rm CS}_{2}=(\y_2 aq^{-1} \T_2 \T_1-1)~.
  \end{split}
\end{equation}
The key observation is that these operators  act on the partition function in a special way. They ''intertwine'' the square  of the $U(1)$ partition function with the $U(2)$ partition function, namely:
\begin{equation}\label{eq:splitCS}
\begin{split}
  &\hat{\mathcal{O}}^{\rm CS}_{1}  \sfZ_{U(2)}^{\rm CS}[y_1,y_2]= -  \sfZ_{U(1)}^{\rm CS}[y_1, aq^{-1}]\sfZ_{U(1)}^{\rm CS}[y_2, a]~,
            \\
            &  \hat{\mathcal{O}}^{\rm CS}_{2}  \sfZ_{U(2)}^{\rm CS}[y_1,y_2]= -  \sfZ_{U(1)}^{\rm CS}[y_1,a]\sfZ_{U(1)}^{\rm CS}[y_2, a q^{-1}]~.
\end{split}
\end{equation}
 This splitting property can be proved explicitly using superintegrability formulas, which we do in Appendix \ref{sec:AppendixSplittingProof}. Finally, let us prove \eqref{eq:AiCS} for the operator $\hat{A}^{\CS}_2$. Using our notation, we get:
 \begin{equation}
  \begin{split}
        & \hat{A}^{\CS}_2 \sfZ_{U(2)}^{\rm CS}[y_1,y_2]= \hat{\sfA}^{\CS}_{U(1)}[\y_1,\T_1]\hat{\mathcal{O}}^{\rm CS}_{2}  \sfZ_{U(2)}^{\rm CS}[y_1,y_2]  
     \\
     &=- \hat{\sfA}^{\CS}_{U(1)}[\y_1,\T_1] \sfZ_{U(1)}^{\rm CS}[y_1, a]\sfZ_{U(1)}^{\rm CS}[y_2, a q^{-1}]=0
  \end{split}
 \end{equation}
 Exactly the same calculation works for $\hat{A}^{\CS}_3$. Thus we have demonstrated that the $U(2)$ partition function solves the quantum variety equations  \eqref{eq:AiCS}-\eqref{eq:QVCS}. Next, we will show that due to their relation to the recursion operator, this solution is also unique. 

 The matrix model partition function is a solution that is regular at $(y_1,y_2)=(0,0)$. To prove uniqueness of this solution we exploit the relation of operators $\hat{A}_i^{\CS}$ to the recursion operator. All of them look structurally similar, so some identity between them can be expected. A direct calculation shows that the recursion operator can be written as a linear combination of \eqref{eq:QVCS}. Explicitly, we have:
\begin{equation}
\begin{aligned}
& (\y_1 - \y_2)\,\hat{\sfA}^{\rm CS}_{U(2)}
= (q \y_1 + q \y_2)\,\hat{A}_1^{\rm CS}  \\
&\quad - \frac{(1+q)q^3}{a}\,\T_{1}^{-1}\T_{2}^{-1}
\Big(
\hat{A}_1^{\rm CS}
+ \hat{A}_2^{\CS}-\hat{A}_3^{\CS}
\Big).
\end{aligned}
\end{equation}

As we have demonstrated in Section \ref{s:CS-model}, the regular solution to the recursion equation is unique (up to normalization). Therefore, due to the relation above, the regular  solution to the rank-2 quantum variety is unique as well. Note that the recursion operator \eqref{eq:operCSrank2} defines the matrix model just as well as the quantum variety \eqref{eq:QVCS}.
\\

Up to this moment we have discussed only solutions which are well behaved around zero. However, the system \eqref{eq:QVCS} has more solutions, which, however, are not well defined around zero, hence do not correspond to the matrix model. We will only discuss this for the $\CS$ model and will not try to provide a full classification.  We would like to mainly address the specific choice that follows directly from the factorized nature of \eqref{eq:QVCS}. Looking at these equations, the most natural approach would be to look for solutions
 $\Psi^{\CS}[y_1,y_2]$ that are simultaneously annihilated by the operators \eqref{eq:CSoperO}
\begin{equation}\label{eq:nonregsolCS}
    \begin{split}
         \hat{\mathcal{O}}^{\rm CS}_{2}  \Psi^{\CS}[y_1,y_2]=0~,
         \\
          \hat{\mathcal{O}}^{\rm CS}_{1}  \Psi^{\CS}[y_1,y_2]=0~.
    \end{split}
\end{equation}
It is a matter of algebra to check that such a function is automatically annihilated by the remaining operator $\hat{A}^{\rm CS}_1$. It follows that for such functions we would have
\begin{equation}
    \hat{y}_1 \Psi^{\CS}[y_1,y_2]= \hat{y}_2 \Psi^{\CS}[y_1,y_2]~.
\end{equation}
This is only possible if the functions have a $\delta$-function like behaviour around $y_1=y_2$. The equation can be solved in Fourier space (with $y_i=\exp u_i$):
\begin{equation}
    \Psi_{[2]}^{\rm CS}[y_1,y_2]= \int dk_1 dk_2 \exp\left(  \dfrac{k_1  \log y_1}{\hbar}+\dfrac{k_2 \log y_2}{\hbar}\right) \Psi_{[2]}^{\rm CS}[k_1,k_2]~.
\end{equation}
After substituting the Fourier transform into the equations we find:
\begin{equation}\label{eq:CSothersol}
    \Psi_{[2]}^{\rm CS}[k_1,k_2]=\exp \left(\left(\frac{1}{2}-N \right)
   \left(k_1+k_2\right) - \dfrac{1}{2\hbar}\left(k_1+k_2 \right)^2\right)
\end{equation}
By this comment we would like to stress that these solutions can be explicitly obtained; however, they are beyond the scope of this paper, as they do not correspond to the matrix model in any obvious way (see \cite{Aganagic:2013jpa} for discussions of similar solutions in topological strings).



Having discussed the quantum operators, now we proceed to the classical limit, which amounts to taking $q \rightarrow 1$ and substituting $\y_i$ with $y_i$ and $\T_i$ with $T_i$. As in the rank 1 case we would like to think of these $q$-difference equations as a quantization of some classical variety. If we were to apply the classical limit to the recursion operator we would get:
\begin{equation}\label{eq:CSAU2cl}
 \begin{split}
   & (\y_1 - \y_2) \hat{\sfA}^{\rm CS}_{U(2)} \frac{1}{(\y_1-\y_2)} \rightarrow\\
    &  \rightarrow\Big ( - 1 - y_1 + T_1^{-1} + a y_1 T_1 \Big ) +  \Big ( - 1 - y_2 + T_2^{-1} + a y_2 T_2 \Big ) = \\
    &\mathcal{A}^{\CS}(y_1,T_1)+\mathcal{A}^{\CS}(y_2,T_2)=0~.
           \end{split}
 \end{equation}
This is not a Lagrangian variety and $\hat{\sfA}^{\rm CS}_{U(2)}$ is not expected to be recovered as its quantization. On the other hand the equations for the quantum variety  \eqref{eq:QVCS} turn into
\begin{equation}\label{CS-C4-variety}
    \begin{split}
&\mathcal{A}^{\CS}(y_1,T_1)=\mathcal{A}^{\CS}(y_2,T_2)~,
    \\
    &\mathcal{A}^{\CS}(y_1,T_1)(a y_2T_1T_2-1)=0~,
    \\
    &\mathcal{A}^{\CS}(y_2,T_2)(a y_1T_1T_2-1)=0~.
    \end{split}
\end{equation}
The three equations in $(\mathbb{C}^*)^4$ define the Lagrangian variety
 with respect to the symplectic structure
 \begin{equation}
     \omega_1 + \omega_2 = \frac{dT_1}{T_1}\wedge \frac{dy_1}{y_1} + 
     \frac{dT_2}{T_2}\wedge \frac{dy_2}{y_2}
 \end{equation}
 and this variety has two components
 \begin{equation}
     {\cal L}_{U(2)}^{\rm CS} = (\Sigma_{a}^{\rm CS} \times \Sigma_{a}^{\rm CS})
     \cup \Gamma_\iota~.
 \end{equation}
  The first component is simply $\Sigma_{a}^{\rm CS} \times \Sigma_{a}^{\rm CS}$
\begin{equation}\label{eq:sigmasigmaCS}
  \Sigma_{a}^{\rm CS} \times \Sigma_{a}^{\rm CS} = \left\{ \begin{split}
        &\mathcal{A}^{\CS}(y_1,T_1) =0
        \\
          &\mathcal{A}^{\CS}(y_2,T_2) =0
  \end{split} \right\}~,
\end{equation}
 and this is Lagrangian in $(\mathbb{C}^*)^4$
 since $\Sigma_{a}^{\rm CS}$ is Lagrangian in $(\mathbb{C}^*)^2$. 
The second component is  defined by the following equations
\begin{equation}\label{eq:secondCS}
 \Gamma_\iota  =  \left\{ \begin{split}
          y_1=y_2
  \, ,\quad  ay_1  T_1T_2-1=0
  \end{split} \right\}~,
\end{equation}
 which guarantee that the original conditions \eqref{CS-C4-variety} are satisfied. This component is actually the graph of the anti-symplectic involution 
 $\iota$ defined in \eqref{CS-def-involution} previously
 \begin{equation}
     \Gamma_\iota = \Big \{  \Big (y_1, T_1, \iota (y_1, T_1)\Big ) = \Big (y_1, T_1, y_1, \frac{1}{ay_1 T_1} \Big ) \in (\mathbb{C}^*)^4 \Big \}~.
 \end{equation}
  The fact that $\iota$ is an anti-symplectic involution in $(\mathbb{C}^*)^2$
   implies that the graph $\Gamma_\iota$ is Lagrangian in $(\mathbb{C}^*)^4$. 
  On $(\mathbb{C}^*)^4$ we have the action which exchanges two $(\mathbb{C}^*)^2$ factors
  \begin{equation}
      \tau (y_1, T_1, y_2 , T_2) = (y_2, T_2, y_1, T_1)~.
  \end{equation}
   The involution property $\iota^2 = {\rm id}$ implies 
   \begin{equation}
       \tau \Big ( \Gamma_\iota \Big ) = \Gamma_{\iota^{-1}} = \Gamma_\iota~.
   \end{equation}

The two components intersect along a diagonal:
\begin{equation}
        y_1=y_2 \,,  \quad  \mathcal{A}^{\CS}(y_1,T_1)=0 \,,\quad \mathcal{A}^{\CS}(y_1,T_2)=0~.
\end{equation}
The topological string interpretation of the two components is explained in great detail in \cite{Aganagic:2013jpa} and \cite{Ekholm:2024ceb}.

As we will discuss in Section \ref{s:semiclassics}, it is the first component \eqref{eq:sigmasigmaCS} that controls the leading term in the semiclassical expansion of the matrix model in the rank-two case. The second component, \eqref{eq:secondCS}, should instead correspond to the leading term in the semiclassical expansion of the full quantum solution \eqref{eq:CSothersol}, which does not appear to admit a direct interpretation in terms of the matrix model. Nevertheless, the full quantum system \eqref{eq:QVCS} is essential once one goes beyond leading order. In particular, subleading terms in the semiclassical expansion of the matrix model receive contributions from both components of ${\cal L}_{U(2)}^{\rm CS}$. We will illustrate this in somewhat greater detail in Section \ref{s:semiclassics}.

\subsection{$q$-Laguerre model.}
Now let us proceed to the $q$-Laguerre matrix model. In this case there are only slight modifications compared to the Chern-Simons model.
We find  three operators that annihilate the partition function \eqref{qL-M2-PF}
\begin{equation}\label{eq-qL-full-pre}
    \hat{A}^{\rm qL}_i \sfZ^{\rm qL}_{U(2)}[y_1,y_2] = 0 \ , \ i=1,2,3~,
\end{equation}
where
\begin{equation}\label{eq-qL-full}
 \begin{split}
        \hat{A}^{\rm qL}_1&= \hat{\sfA}^{\rm qL}_{U(1)}[\y_1,\T_1] - \hat{\sfA}^{\rm qL}_{U(1)}[\y_2,\T_2]~,
        \\\hat{A}^{\rm qL}_2 &=\hat{\sfA}^{\rm qL}_{U(1)}[\y_1,\T_1] \left(\y_2( - a^2 q^{-1}\T_1 \T_2+1) -1\right)~,
           \\
              \hat{A}^{\rm qL}_3&= \hat{\sfA}^{\rm qL}_{U(1)}[\y_2,\T_2] \left(\y_1( - a^2 q^{-1}\T_1 \T_2+1) -1\right)~.
 \end{split}
\end{equation}
We observe that the system is structurally similar to the $\CS$ one. Denote the operators that appear on the right as
\begin{equation}
    \hat{\mathcal{O}}^{\qL}_i=\left(\y_i( - a^2 q^{-1}\T_1 \T_2+1) -1\right)\,.
\end{equation}
They play the same role of separating the variables, via equations
\begin{equation}
    \hat{\mathcal{O}}^{\rm qL}_{2} \sfZ^{\rm qL}_{U(2)}[y_1,y_2]=-\sfZ^{\rm qL}_{U(1)}[y_1, a]\sfZ^{\rm qL}_{U(1)}[qy_2, aq^{-1}]~,
\end{equation}
which implies that the rank 2 partition function is annihilated by operators $ \hat{A}^{\rm qL}_i$.  Uniqueness of the solution is proved by relating the system to the rank 2 recursion operator. We have the following operator relation
\begin{equation}
    \begin{split}
& (1-a^2 q^{-2} \, \T_{2} \T_{1} )
\Big[ (\y_1 - \y_2)\, \hat{\sfA}^{\rm qL}_{U(2)} 
- (\y_1 + \y_2)\, \hat{A}^{\rm qL}_1 \Big]+  \\
&\quad + (1+q)\Big[
\hat{A}^{\rm qL}_1 
+ \hat{A}^{\rm qL}_2 
- \hat{A}^{\rm qL}_3 
\Big]=0~.        
    \end{split}
\end{equation}

The system \eqref{eq-qL-full} also has non-regular solutions of type \eqref{eq:nonregsolCS}, which we will not explore here.
\\

The classical limit defines a Lagrangian variety in $(\mathbb{C}^*)^4$
\begin{equation}
    {\cal L}_{U(2)}^{\qL}= \left\{\begin{gathered}
    (y_1,y_2,T_1,T_2) \in (\mathbb{C}^*)^4
      \\
        \mathcal{A}^{\rm qL}(y_1,T_1)=  \mathcal{A}^{\rm qL}(y_2,T_2)
        \\
           \mathcal{A}^{\rm qL}(y_1,T_1)\left(y_2( - a^2 T_1 T_2+1)-1\right) = 0\\  \mathcal{A}^{\rm qL}(y_2,T_2)\left(y_1( - a^2 T_1 T_2+1)-1\right) = 0
    \end{gathered} \right\}
\end{equation}
which has two components
\begin{equation}
     {\cal L}_{U(2)}^{\rm qL} = (\Sigma_{a}^{\rm qL} \times \Sigma_{a}^{\rm qL})
     \cup \Gamma_\iota 
\end{equation}
 such that 
\begin{equation}
    \begin{gathered}
 \Sigma^{\qL}_a \times \Sigma^{\qL}_a = \left \{ \mathcal{A}^{\rm qL}(y_1,T_1)=0~,~~\mathcal{A}^{\rm qL}(y_2,T_2)=0 \right \}~,
      \\
 \Gamma_\iota= \left\{   y_1=y_2 \quad  \left(y_1( - a^2 T_1 T_2+1)-1\right) = 0 \right\}~.
    \end{gathered}
\end{equation}
 The first component is Lagrangian in $(\mathbb{C}^*)^4$ since $\Sigma^{\qL}_a$ is Lagrangian in $(\mathbb{C}^*)^2$. The second component
  $\Gamma_\iota$ is the graph of the birational anti-symplectic involution 
   defined in \eqref{qL-involution-def}
   \begin{equation}
     \Gamma_\iota = \Big \{  \Big (y_1, T_1, \iota (y_1, T_1)\Big ) = \Big (y_1, T_1, y_1, \frac{y_1-1}{a^2 y_1 T_1} \Big ) \in (\mathbb{C}^*)^4 \Big \}~.
   \end{equation}
  Thus $\Gamma_\iota$ is a Lagrangian submanifold in $(\mathbb{C}^*)^4$.

\subsection{$q$-Gaussian model}

In the $q$-Gaussian matrix model the rank 1 curve is of degree two in $y$. Unexpectedly this causes an additional complication in the computation of the rank two quantum variety. Recall that
the quantum curve can be rewritten as
 \begin{equation}
     q^{-2}\T \hat{\sfA}^{\rm qG}_{U(1)} = (1-\T)-\y^2(1-a \T)(1-q a\T)~.
 \end{equation}
 The quantum difference equations for the rank 2 $q$-Gaussian model turn out to be somewhat more complicated and we found two ways of writing them. The system will consist of three equations as before. The first equation is the expected
 \begin{equation}
     \hat{A}^{\rm qG}_1 \sfZ_{U(2)}^{\rm qG}[y_1,y_2]=0
 \end{equation}
 with
 \begin{equation}
    \hat{A}^{\rm qG}_1 = \hat{\sfA}_{U(1)}^{\rm qG}[\y_1, \T_1]- \hat{\sf{A}}_{U(1)}^{\rm qG}[\y_2, \T_2]\,.
\end{equation}
This operator annihilates the partition function for the same reason as in the Chern-Simons and $q$-Laguerre cases. 
 The function $\sfZ_{U(2)}^{\rm qG}[y_1,y_2]$ is defined in \eqref{def-qG-M2}. 

To find the other equation we follow the idea of the splitting property that we used above. However, this time it works in a somewhat subtle way. The operator that plays the role of the splitting operator for the $q$-Gaussian matrix model is
\begin{equation}
     \hat{\mathcal{O}}^{\qG}_2= \y_2^2 (1-a^2 q^{-1}\T_1 \T_2)-1
\end{equation}
and $ \hat{\mathcal{O}}^{\qG}_1$ with the labels $1$ and $2$ interchanged. Instead of complete separation of variables as a result of its action we get a sum of two components with each of them being a product of two $U(1)$ partition functions:
\begin{equation}\label{eq:splitgaus}
   \begin{split}
        \hat{\mathcal{O}}^{\qG}_2 \sfZ^{\rm qG}_{U(2)}[y_1,y_2] &=\dfrac{a-1}{a q} y_1 y_2\qgPF{y_1,aq} \qgPF{q y_2,aq^{-1}} -
        \\
        &-\qgPF{y_1,a} \qgPF{q y_2,aq^{-2}} ~.
   \end{split}
\end{equation}
This separation seems to have something to do with the additional parity symmetry of the $q$-Gaussian model. Namely, the rank two partition function is invariant only under    simultaneous reflection of $y_1,y_2$
  \begin{equation}
       \sfZ^{\qG}_{U(2)}[-y_1, -y_2] =  \sfZ^{\qG}_{U(2)}[y_1, y_2]~.
  \end{equation}
At the same time, a product of two rank one partition functions is invariant under reflections of $y_1$ and $y_2$ separately and so is the splitting operator. This apparent mismatch is reflected in the structure of eq. \eqref{eq:splitgaus}. 

Due to this effect we cannot simply act with the rank one quantum curve from the left to obtain $\hat{A}^{\qG}_2$ and $\hat{A}^{\qG}_3$. However, a slightly modified version of the same argument can still be applied. We define:
 \begin{equation}\label{eq:qgA2}
    \begin{split}
   \hat{A}_2^{\rm qG}=     \Big(&aq^{-2}  \y_1^2\y_2(1- a q^{-2}\T_2)(1-a q\T_{1})    \hat{\sfA}_{U(1)}^{\qG}[\y_1,\T_1,aq] \hat{y}_1^{-1} +
        \\+&{(1-q^{-2}a)} \hat{\sfA}_{U(1)}^{\rm qG}[\y_1,\T_1,a] \Big) \cdot \hat{\mathcal{O}}^{\qG}_2  
    \end{split}
\end{equation}
and $ \hat{A}_3^{\rm qG}$, which is obtained by exchanging the $1$ and $2$ indices. We claim that such operators annihilate the rank two partition function
\begin{equation}\label{eq:A2A3qg}
   \begin{split}
       \hat{A}_2^{\rm qG} \sfZ_{U(2)}^{\rm qG}[y_1,y_2] = 0~,
       \\
       \hat{A}_3^{\rm qG} \sfZ_{U(2)}^{\rm qG}[y_1,y_2] = 0~.
   \end{split}
\end{equation}
The equations \eqref{eq:A2A3qg} work as follows. When we act with the left part of \eqref{eq:qgA2} on the image of $\hat{\mathcal{O}}^{\qG}_2 $ the operators $\hat{\sfA}_{U(1)}^{\rm qG}[\y_1,\T_1,a]$ and $\hat{\sfA}_{U(1)}^{\qG}[\y_1,\T_1,aq] \hat{y}_1^{-1}$ annihilate their counterparts in \eqref{eq:splitgaus}, i.e.:
\begin{equation}
    \hat{\sfA}_{U(1)}^{\qG}[\y_1, \T_1, aq] \hat{y}_1^{-1}  \cdot y_1  \qgPF{y_1,aq} =0 \,,\quad  \hat{\sfA}_{U(1)}^{\rm qG}[\y_1,\T_1,a] \cdot \qgPF{y_1,a}=0~.
\end{equation}
The cross terms do not vanish individually, but cancel after applying several identities. First we use
\begin{equation}\label{eq:AZidentity1}
    \y_1 \hat{\sfA}_{U(1)}^{\qG} [\y_1, \T_1, aq] \y_1^{-1} \sfZ_{U(1)}^{\qG}[y_1, a]= q^2 (q-1) \sfZ_{U(1)}^{\qG}[y_1,a]
\end{equation}
and 

\begin{equation}\label{eq:AZidentity2}
    \hat{\sfA}_{U(1)}^{\qG} [\y_1, \T_1, a] \y_1 \sfZ_{U(1)}^{\qG}[y_1, aq]= -q (q-1) y_1\sfZ_{U(1)}^{\qG}[y_1, aq]
\end{equation}
 Finally, the operator $(1- a q^{-2}\T_2)(1-a q\T_{1})$ allows us to relate the two terms, by using identities like \eqref{eq:contiguousappendix} and \eqref{eq:contiguousappendix2}  for Chern-Simons theory.

Note that there seems to be another possibility, which we briefly mention. Note that after applying \eqref{eq:AZidentity1} and \eqref{eq:AZidentity2} we could keep acting with  $ \hat{\sfA}_{U(1)}^{\qG} [\y_1, \T_1]$ from the left. Then we would have:
 \begin{align}
 & \hat{\sfA}_{U(1)}^{\qG} [\y_1, \T_1,a]  \y_1   \hat{\sfA}_{U(1)}^{\qG} [\y_1, \T_1,a q]  \y_1^{-1} \hat{\mathcal{O}}^{\qG}_2  \cdot \sfZ^{\rm qG}_{U(2)}[y_1,y_2] \\
 & = \hat{\sfA}_{U(1)}^{\qG} [\y_1, \T_1, aq]  \y_1^{-1}   \hat{\sfA}_{U(1)}^{\qG} [\y_1, \T_1,a]  \y_1 \hat{\mathcal{O}}^{\qG}_2  \cdot \sfZ^{\rm qG}_{U(2)}[y_1,y_2]=0
 \end{align}
Therefore we could set:
\begin{equation}
    \hat{B}_2^{\qG}= \hat{\sfA}_{U(1)}^{\qG} [\y_1, \T_1,a]  \y_1   \hat{\sfA}_{U(1)}^{\qG} [\y_1, \T_1,a q]  \y_1^{-1} \hat{\mathcal{O}}^{\qG}_2 
\end{equation}
and $ \hat{B}_3^{\qG}$ with $(y_1,T_1)$ and $(y_2, T_2)$ interchanged. However, as we will see below, such operators cannot be obtained by quantizing a classical variety, therefore, we mainly stick with \eqref{eq:A2A3qg}.
\\

This choice of $q$-difference operators $\hat A_{1,2,3}^{\qG}$ defines in the classical limit
the Lagrangian variety ${\cal L}_{U(2)}^{\rm qG}$ in $(\mathbb{C}^*)^4$
\begin{equation}\label{qG-full-set}
  \begin{split}
       & \hat A_1^{\qG} \rightarrow \mathcal{A}^{\qG}(y_1,T_1)= \mathcal{A}^{\qG}(y_2,T_2)~,\\
       & \hat A_2^{\qG} \rightarrow \begin{split}
           \Big(a y_1 y_2& (1-a T_1)(1-a T_2)+(1-a) \Big)\times\\
           &\times\mathcal{A}^{\qG}(y_1,T_1)\left( y_2^2 (1-a^2 T_1 T_2)-1\right)=0~,
       \end{split}\\
       & \hat A_3^{\qG} \rightarrow 
       \begin{split}
                  \Big(a y_1 y_2& (1-a T_1)(1-a T_2)+(1-a) \Big)\times
                  \\&\times  \mathcal{A}^{\qG}(y_2,T_2)\left( y_1^2 (1-a^2 T_1 T_2)-1\right)=0~,
       \end{split}
  \end{split}
\end{equation}
 which has 4 components
\begin{equation}
   {\cal L}_{U(2)}^{\rm qG} = (\Sigma^{\rm qG}_a \times \Sigma^{\rm qG}_a) \cup \Gamma_{\iota_+} \cup \Gamma_{\iota_-} \cup \Gamma_{\iota_s}~. 
\end{equation}
 The first component is defined as 
 \begin{equation}
 \Sigma^{\qG}_a \times \Sigma^{\qG}_a = \{ \mathcal{A}^{\qG}(y_1,T_1)=0~,~~\mathcal{A}^{\qG}(y_2,T_2)=0 \}~,
\end{equation}
 which is Lagrangian in $(\mathbb{C}^*)^4$ since $ \Sigma^{\qG}_a$ is Lagrangian 
  in $(\mathbb{C}^*)^2$. Two other components are defined as follows
\begin{equation}
 \Gamma_{\iota_\pm} = \left\{   y_1= \pm y_2~,~~  \left( y_1^2 (1-a^2 T_1 T_2)-1\right) = 0 \right\}
\end{equation} 
 and they correspond to the graphs of birational anti-symplectic involutions 
 $\iota_\pm$ defined in \eqref{qG-involution-def}
 \begin{equation}
      \Gamma_{\iota_\pm} = \Big \{ \Big ( y_1, T_1, \iota_\pm (y_1, T_1 ) \Big ) \Big \}
 \end{equation}
  and thus these two components are Lagrangian. The last component
 corresponds to the graph of the birational anti-symplectic involution $\iota_s$ defined in \eqref{iotas-def}-\eqref{eq-qG-invol-extra2}
 \begin{equation}
   \Gamma_{\iota_s} = \Big \{ \Big ( y_1, T_1, \iota_s (y_1, T_1 ) \Big ) \Big \}~.
 \end{equation}
  This is Lagrangian in $(\mathbb{C}^*)^4$ and it satisfies the properties
  \eqref{ext-invol-prop1} and \eqref{ext-invol-prop2} and thus it satisfies 
   the conditions \eqref{qG-full-set}. At the moment we do not have any clear geometrical argument why the involution $\iota_s$ plays a role in this construction.

 \subsection{Comment on higher rank}

 We would like to interpret $\sfZ_{U(M)}[y_1,\ldots,y_M]$ as the quantization of a Lagrangian variety in $(\mathbb{C}^*)^{2M}$. From this perspective, one expects the partition function to be uniquely characterized by a system of $q$-difference equations whose classical limit defines the corresponding Lagrangian variety. Constructing such a system explicitly, however, turns out to be technically rather involved.

Following the expectations of \cite{Aganagic:2013jpa}, we have checked that, at rank $M=3$ and for all the models considered here, the analogue of the $\hat{A}_1$ constraint is given by the completely antisymmetric product
 \begin{equation}
    \prod_{1\leq i<j\leq 3}\left( \hat{\sfA}_{U(1)}^{w}[\y_i, \T_i] - \hat{\sfA}_{U(1)}^{w}[\y_j, \T_j]\right)\cdot \sfZ_{U(3)}^{w}[y_1,y_2,y_3]=0~.
\end{equation}
At present, however, we do not know how to derive the remaining independent quantum constraints.

\section{Semiclassical expansion}\label{s:semiclassics}

In this section we discuss the semiclassical expansion of matrix model partition functions.
The semiclassical limit of the matrix model is the regime when
\begin{equation}
    \hbar=\log q \rightarrow0 \qquad N \rightarrow \infty~,
\end{equation}
such that $N\hbar$ is fixed, in other words, the variable $a=q^{N}$ remains finite. The conjecture is that the semiclassical expansion of the partition function assumes a WKB type form
\begin{equation}
    \sfZ(\bp)= \exp\left( \dfrac{1}{\hbar} \sfF(\bp,\hbar,a) \right)~,
\end{equation}
where $\sfF$ is given as:
\begin{equation}
   \sfF(\bp,\hbar,a) = \sum_{n=0}^{\infty} \hbar^{n}\sfF_{n}(\bp,a)~.
\end{equation}
We can use the $q$-difference equations for partition functions to compute terms in this expansion. The fact that our equations can be understood as quantized Lagrangian varieties guarantees the validity of the semiclassical ansatz. We will describe the general procedure first and discuss combinatorics relevant to different models below.

Starting with rank one, we set $p_k=y^k$ and consider the quantum curve equation. In the leading order we simply get  the classical curve
\begin{equation}
    \mathcal{A}(y,T)=0~,
\end{equation}
where we identify $T= \exp\left( y\dfrac{d}{dy}\sfF_0[y,a]\right)$. In the examples at hand, the equations are quadratic in $T$, hence there are two possible branches $T_{\pm}$.  Only one of these branches has the correct behaviour at $y \rightarrow 0$ and   corresponds to the matrix model.

We could use this equation to expand further and find corrections to the leading order partition function in rank one. However, beyond leading order we have contributions which come from ''multi-trace'' correlators. 

At rank two the full partition function is described by the quantum equations $\hat{A}_i^{w}$. Therefore the leading order will be determined by the equation of the classical variety with the identification:
\begin{equation}
    T_i = \exp\left( y_i\dfrac{\partial}{\partial y_i } \sfF_0 [y_1,y_2,a]\right)\,,\quad i=1,2~.
\end{equation} 
The classical rank two variety has several components. However, the matrix model requires the solution $T_i$ to be regular as $y_i \rightarrow 0$, which means that the solution lies on the $\Sigma_{U(1)} \times \Sigma_{U(1)}$ component of the variety. Hence we find that:
\begin{equation}
    \sfF_0[y_1,y_2,a]=\sfF_{0}[y_1,a]+\sfF_{0}[y_2,a]
\end{equation}
and 
\begin{equation}
    T_i = \exp\left( y_i\dfrac{\partial}{\partial y_i } \sfF_0 [y_i,a]\right)\,,\quad i=1,2
\end{equation}
Such a factorization is expected from the matrix model side and predicts that the leading order term has the following structure at generic rank:
\begin{equation}
    \sfF_0(\bp,a)=\sum_{k=1}^{\infty} p_k \Phi_{[k],0}(a)
\end{equation}
This implies that $\Phi_{[k],0}(a)$ is the leading order contribution to ''single trace'' correlators, which we will confirm by explicit computations below.

When going to next-to-leading order, it is crucial to keep in mind that $ \sfF_1[y_1,y_2|a]$ does not split into a sum of independent contributions for each $y_i$. This can be checked explicitly using superintegrability formulas. Hence beyond leading order, the matrix model gets quantum corrections from the second component of the quantum variety as well as from the first component. This demonstrates once again that quantizing the $\Sigma_{U(1)} \times \Sigma_{U(1)}$ component only would not produce a correct description of the matrix model.
In all our examples we find the following expansion
\begin{equation}\label{eq:genus1}
    \begin{split}
        \sfF_1[y_1,y_2,a] =& \sum_{k=1}^{\infty} (y_1^{k}+y_2^{k}) \Phi_{[k],1}(a)+
        \\+&\sum_{k_1 \geq k_2=1}^{\infty} (y_1^{k_1}+y_2^{k_1})(y_1^{k_2}+y_2^{k_2}) \Phi_{[k_1,k_2],1}(a) ~.
    \end{split}
\end{equation}
We will demonstrate how this works below. Note that this means that both  $\Phi_{[k],1}(a)$ and $\Phi_{[k_1,k_2],1}(a)$ get contributions from the $\Gamma_{\iota_k}$ components. These contributions are separated in such a way that $\Phi_{[k],1}(a)$ should correspond to the subleading order of ''single-trace'' correlators and $\Phi_{[k_1,k_2],1}(a)$ to the leading order of connected ''double-trace'' correlators. The fact that such a contribution only appears at the next-to-leading order in the full partition function is a non-trivial observation. It suggests that the semiclassical expansion has a genus-like structure. As a sidenote, we mention that, based on the leading and subleading terms it is reasonable to conjecture that at generic rank the expansion has the form:
\begin{equation}
    \sfF_{n}(\bp,a)  =
    \sum_{l=1}^{n+1}\sum_{k_1,\ldots, k_l \geq 1} p_{[k_1,\ldots,k_l ]}\Phi_{[k_1,\ldots,k_l ],n}(a)~.
\end{equation}
where $p_{[k_1,\ldots,k_l ]}=\prod\limits_{j=1}^{l}p_{k_j}$. Such a structure implies that after the time variables are rescaled as $p_k = \hbar\,\bar{p}_k$ one obtains a genus expansion:
\begin{equation}
    \log \sfZ(\bar{\bp}) = \sum_{g=0}^{\infty}\sum_{n=1}^{\infty} \hbar^{2g-2+n} \sum_{k_1,\ldots, k_n \geq 1} \bar{p}_{[k_1,\ldots,k_n ]}\Phi_{[k_1,\ldots,k_n ],g}(a)~.
\end{equation}

We will mostly discuss explicit formulas for the leading-order contributions. The validity of the ansatz will be discussed elsewhere.

\subsection{Leading order and explicit combinatorics.}

The leading semiclassical partition function determines the behaviour of ''single trace'' correlators. Let us denote these as
\begin{equation}
    m_n^w(q,a)=\ev{\sum\limits_{i=1}^N x_i^n }^w 
\end{equation}
and their generating function as
\begin{equation}
    \mathfrak{m}^w(y|q,a) = \sum_{n=0}^{\infty} y^n m_n^w(q,a)~.
\end{equation}
These moments can be calculated explicitly using various techniques \cite{forrester2023q,byun2024q}, including superintegrability. Denote the semiclassical limits as 
\begin{equation}\label{eq:limitsingletrace}
\begin{split}
       & \lim_{\substack{\hbar \rightarrow 0\\
    N \rightarrow\infty}} \hbar \, m_n^w(q,a) = \mu_n^{w}(a)~,
       \\
       & \lim_{\substack{\hbar \rightarrow 0\\
    N \rightarrow\infty}} \hbar \,\mathfrak{m}^w(y|q,a) = \upmu^{w}(y|a)~.
\end{split}
\end{equation}
The asymptotic expressions $\upmu^{w}(y|a)$ were studied in \cite{forrester2023q,byun2024q}, using exact expressions for the moments $  m_n^w(q,a)$ of $q$-deformed matrix models. Below we will reproduce these results case by case by solving the corresponding classical curves.  Along the way we note some extra combinatorial structures that generalize the corresponding undeformed matrix model combinatorics.

If the semiclassical expansion captures the behaviour of correlation functions correctly, we should identify
\begin{equation}
     \upmu^{w}(y|a)=  y\dfrac{d}{dy } \sfF^w_0(y|a) ~.
\end{equation}
Hence the classical curve becomes:
\begin{equation}
    \mathcal{A}^w(y, e^{\upmu^{w}(y|a)})=0~.
\end{equation}
The equation is then solved straightforwardly. 

\paragraph{Chern-Simons model.}

From the classical curve we get:
\begin{equation}\label{eq:CSTpm}
    T^{\pm}=\frac{1+y \pm\sqrt{-4 a y+y^2+2 y+1}}{2 a y}
\end{equation}
Only the $T^{-}$ branch corresponds to the matrix model. This gives
\begin{equation}
    \upmu^{\rm CS}(y|a)=y\dfrac{d}{dy}\log\left(\frac{1+y-\sqrt{-4 a y+y^2+2 y+1}}{2 a y} \right)=\sum_{n=1}^{\infty} y^n \mu^{\rm CS}_n(a)~.
\end{equation}
For the individual moments we have:
\begin{equation}
    \mu^{\rm CS}_n(a) = \dfrac{(a-1)}{n}\sum_{k=0}^{n-1} (-1)^{k+n-1} \binom{n-1}{k}\binom{n+k}{n} a^k
\end{equation}
This reproduces the result of \cite{forrester2022global}. Further, we see that apart from the prefactor, the quantity is finite at $a=1$, namely:
\begin{equation}
   \dfrac{ \mu^{\rm CS}_n(a)}{a-1} \Bigg|_{a=1 }=1
\end{equation}

\paragraph{The $q$-Gaussian model.}

The classical curve (in a convenient normalization) is given by:
\begin{equation}
    \mathcal{A}^{\rm qG}(y,T)=1-T-y^2(1- aT)^2
\end{equation}
which gives:
\begin{equation}
    \upmu^{\rm qG}(y|a)=\log \left(\frac{2 a y^2-1+\sqrt{1+4 a y^2(a-1)}}{2 a^2 y^2}\right) 
\end{equation}
The individual moments reproduce the result obtained in \cite{byun2024q}, namely:
\begin{equation}\label{eq:qgmoments2}
    \mu_{2n}^{\rm qG}(a)=-\dfrac{1}{n}\left(1+\dfrac{(2n)!}{(n-1)!}\sum_{k=0}^{n} \dfrac{(-1)^{k+1}a^{k+n}}{(k+n)k!(n-k)!} \right)
\end{equation}
This formula can be simplified to elucidate an important structure:
\begin{equation}\label{eq:qgmoments3}
   \mu_{2n}^{\rm qG}(a)=(-1)^{n} (a-1)^{n+1} \left( \dfrac{1}{n} \sum_{k=0}^{n-1} \binom{n+k}{n} a^k\right)
\end{equation}
We expect that in the direct $\hbar \rightarrow 0$ limit, we also have $a \rightarrow1$ and the result should reproduce the moments of the classical Gaussian matrix models. The representation \eqref{eq:qgmoments3} suggests that the properly normalized quantity is:
\begin{equation}
   C_n(a)= \dfrac{\mu_{2n}^{\rm qG}(a)}{(-1)^{n} (a-1)^{n+1}}=\left( \dfrac{1}{n} \sum_{k=0}^{n-1} \binom{n+k}{n} a^k\right)
\end{equation}
This representation  of the leading order  correlator makes the relation to the classical Gaussian matrix model explicit. The polynomials $C_n(a)$ can be viewed as an explicit deformation of Catalan numbers. Namely, at $a=1$ we get the Catalan number:
\begin{equation}
    C_n(1)=C_n = \dfrac{1}{n+1}\binom{2n}{n}
\end{equation}
The appearance of Catalan numbers (and their higher genus counterparts) in the large $N$ expansion of the Gaussian matrix model can be seen as a consequence of the graph expansion. While the full $q$-deformed analogue of the combinatorics of the $q$-Gaussian matrix models and their potential relation to some version of $q$-Catalan numbers is yet to be uncovered, we see that a certain natural $a$-dependent quantity appears in the semiclassical limit.

\paragraph{The $q$-Laguerre model.}
The $q$-Laguerre model doesn't lead to any new combinatorics. In fact, exactly like in the undeformed case, the leading-order is essentially the same as in the $q$-Gaussian model. Namely, the generating function of the leading order moments only differs by substituting $y^2$ for $y$:
\begin{equation}
    \upmu^{\qL}(y|a)=\log \left(\dfrac{-1+2ay+\sqrt{1+4 a y(a-1)}}{2 a y} \right)
\end{equation}

\subsubsection{Leading order density}

The rank one partition function in the semiclassical limit is nothing but the matrix model resolvent. Therefore we can also find the leading order eigenvalue density. We will only briefly mention the $q$-Gaussian model. In fact, for all the models in question the leading order densities have been obtained in \cite{forrester2023q,byun2024q,forrester2022global}. In our case we compute the density as the discontinuity of the resolvent. Denote $z=y^2$, then:
\begin{equation}
    \rho_{(0)}^{\rm qG}(z|a) = \upmu(z+i \epsilon|a)- \upmu(z-i \epsilon|a)
\end{equation}
We can compare this to the density found in \cite{byun2024q}. We find the following agreement after rescaling of variables, for $a>1/2$:
\begin{equation}
    \dfrac{i}{2}\rho_{(0)}^{\rm qG}\left(\dfrac{1}{x^2(a-1)}\Bigg|\, a\right) = (\pi\lambda x) \dfrac{1}{1-a}\tilde{\rho}^{\rm dH}_{(0)}\left(xa \sqrt{1-a} \Bigg|\,a\right)
\end{equation}
with the density $\tilde{\rho}^{\rm dH}_{(0)}$ of \cite[eq. (2.27)]{byun2024q},
where we should identify $a=\exp(-\lambda)$. Similarly, the leading-order for the $q$-Laguerre and Chern-Simons model can be found and reproduce the results found in \cite{forrester2023q}.

\subsection{Beyond leading order.}

As we discussed, to go beyond leading order we need higher rank quantum varieties. For rank two we can go to next-to-leading order. Let us illustrate how this works for the Chern-Simons matrix model. We substitute the following ansatz
\begin{equation}
    \sfZ^{\CS}_{U(2)}[y_1,y_2]=\exp\left({\dfrac{1}{\hbar} \left( \sfF_0[y_1,a]+\sfF_0[y_2,a] \right)}+\sfF_1 [y_1,y_2,a] +\ldots \right)~.
\end{equation}
After taking into account the leading-order expression \eqref{eq:CSTpm}, the quantum operators become:
\begin{equation}
    \hat{T}_i \rightarrow T^{-}(y_i)+\hbar \left( T^{-}(y_i)  \upmu_{(1),i}^{\CS}(y_1,y_2)+\dfrac{1}{2}  y_i \dfrac{\partial T^{-}(y_i)}{\partial y_i}\right) ~,
\end{equation}
where $ \upmu_{(1),i}^{\CS}(y_1,y_2) = y_i \frac{\partial}{\partial y_i}\sfF^{\CS}_1[y_1,y_2]$. Due to the classical variety $\mathcal{L}_{U(2)}^{\CS}$ being a Lagrangian variety in $(\mathbb{C}^*)^4$,  at order $\hbar$ we obtain a system of linear equations for $\upmu_{(1),1}^{\CS}(y_1,y_2),\upmu_{(1),2}^{\CS}(y_1,y_2)$, which can be solved purely algebraically. Then one can simply integrate, for example, $\upmu_{(1),1}^{\CS}(y_1,y_2)$ once
\begin{equation}
    \sfF^{\CS}_1[y_1,y_2]= \int\dfrac{dy_1}{y_1} \upmu_{(1),1}^{\CS}(y_1,y_2)+c(y_2)
\end{equation}
and fix the $y_2$ dependence by requiring symmetry under $y_1 \leftrightarrow y_2$. As a result we obtain:
\begin{equation}
\begin{split}
          &\sfF^{\CS}_1[y_1,y_2]=
          \\
          &=\dfrac{1}{2}\log\left( \frac{a^6 \left(a-1+\chi _1\right){}^5 \left(a-1+\chi _2\right){}^5}{\chi _1 \chi
   _2 \left(a(1-a)+\chi _1^2\right) \left(a(1-a)+\chi _1 \chi _2\right){}^2
   \left(a(1-a)+\chi _2^2\right)} \right) =\\
   &=\dfrac{1}{2} \log \left( \dfrac{a^3(a-1+\chi_1)^5}{\chi_1\left(a(1-a)+\chi_1^2\right)}  \right)+\dfrac{1}{2}\log \left( \dfrac{a^3(a-1+\chi_2)^5}{\chi_2\left(a(1-a)+\chi_2^2\right)}  \right)  -
   \\&- \log\left(a(1-a)+\chi_1\chi_2 \right)\Big)~,
\end{split}
\end{equation}
where for compactness we introduced a notation $\chi_i=\dfrac{1-a T^{-}(y_i)}{T^{-}(y_i)}$. Expanding this generating function in $y_1,y_2$ and rearranging the expansion in terms of power sums $p_{k}(\by)=y_1^k+y_2^k$ we find that it follows the pattern \eqref{eq:genus1}. First, let us look at the term linear in $p_k(\by)$. We find that:
\begin{equation}
    \Phi_{[k],1}(a)= \dfrac{k}{2} \Phi_{[k],0}(a)
\end{equation}
which means it can be eliminated by introducing an extra scaling of $q^{k\over 2}$ to the time variables.  In fact, one can check, using the expression for the full quantum moments, that this eliminates all odd order contributions $\Phi_{[k],2n+1}$ \cite{forrester2022global}. From the mixed $y_1,y_2$ terms we get contributions quadratic in power sums, which have the form:
\begin{equation}
\begin{split}
        \Phi_{[k_1,k_2],1}(a)= \dfrac{a(a-1)}{k_1+k_2}\Big(& P^{(1,0)}_{{k_1-1}}(2a-1)P^{(0,1)}_{k_2-1}(2a-1)+\\
    &P^{(1,0)}_{k_2-1}(2a-1)P^{(0,1)}_{k_1-1}(2a-1) \Big) \quad k_1 \geq k_2~,
\end{split}
\end{equation}
where $P^{(1,0)}_{{k-1}}(2a-1)$ and $P^{(0,1)}_{{k-1}}(2a-1)$ are the Jacobi polynomials, given by the expansions:
\begin{equation}
\begin{split}
        P_{k-1}^{(1,0)}(2a-1)=
\sum_{j=0}^{k-1}
(-1)^{k-1+j}
\binom{k-1}{j}
\binom{k+j}{j}
a^j~,\\
 P_{k-1}^{(0,1)}(2a-1)=
\sum_{j=0}^{k-1}
(-1)^{k-1+j}
\frac{k}{j+1}
\binom{k-1}{j}
\binom{k+j}{j}
a^j~.
\end{split}
\end{equation}
This should be matched with ''double-trace'' correlation functions in the matrix model. More precisely, because of the exponentiated structure of the free energy, we should consider connected correlators. These are given by the following expressions:
\begin{equation}
    m_{[k_1,k_2]}^{\CS}(q,a)= \ev{\sum_{i=1}^N x_i^{k_1}\sum_{i=1}^N x_i^{k_2}}^{\CS} -\ev{\sum_{i=1}^N x_i^{k_1}}\ev{\sum_{i=1}^N x_i^{k_2}}^{\CS}
\end{equation}
They can be computed exactly for any $k_1,k_2$ using superintegrability formulas. We checked that after taking the semiclassical limit we have
\begin{equation}
    \lim_{\substack{\hbar \rightarrow 0\\
    N \rightarrow\infty}} m_{[k_1,k_2]}^{\CS}(q,a) = \dfrac{\Phi_{[k_1,k_2],1}(a)}{k_1k_2}
\end{equation}
Note the absence of the extra $\hbar$ factor, compared to the limit of ''single trace'' correlators \eqref{eq:limitsingletrace}. This is consistent with the fact that this contribution in the free energy only appears at subleading order and with the structure of a genus-like expansion. Using the rank-two $q$-difference operators, one can obtain analogous results for other matrix models; however, such an analysis lies beyond the scope of this paper.

\section{Discussion}\label{s:discussion}

In this work, we have analysed in detail three $q$-deformed matrix models: the Chern-Simons, $q$-Laguerre, and $q$-Gaussian models. We have focused on $\sfZ_{U(1)}[y]$, the expectation value of a single inverse characteristic polynomial, and on $\sfZ_{U(2)}[y_1,y_2]$, the expectation value of two inverse characteristic polynomials. Exploiting the superintegrability of these models, we have derived the explicit difference operator $\hat{\sfA}_{U(1)}$ annihilating $\sfZ_{U(1)}[y]$, as well as the operator $\hat{\sfA}_{U(2)}$ annihilating $\sfZ_{U(2)}[y_1,y_2]$.

The function $\sfZ_{U(1)}[y]$ admits a direct interpretation as the quantization of a classical curve obtained by taking the classical limit of $\hat{\sfA}_{U(1)}$. The interpretation of $\sfZ_{U(2)}[y_1,y_2]$ as the quantization of a Lagrangian variety in $(\mathbb{C}^*)^4$ is more subtle. In this case, the single operator $\hat{\sfA}_{U(2)}$ has to be replaced by an appropriate system of quantum constraints. Within our present understanding, the construction of this system remains somewhat ad hoc. For the three matrix models considered here, we were able to identify the required operators by combining insight from open topological strings with guesswork.

One potential implication of our result is that the semiclassical $\hbar \rightarrow 0 ,N\rightarrow \infty$ expansion of the $q$-Gaussian and $q$-Laguerre matrix models admits a topological string description. The resulting free energy expansion depends on $a=q^{N}$, which remains finite in this limit. On the other hand taking the direct $q \rightarrow 1$ limit at fixed $N$ should reproduce the Gaussian matrix model, which then admits its own large $N$ expansion. We see in examples that taking the semiclassical limit first and then taking $a \rightarrow 1$ reproduces the large $N$ expansion of the classical matrix model at leading order and even at the next-to-leading order (see also \cite{byun2024q}). However, a general understanding and a geometric picture behind the commutativity of these limits remain to be understood.


A natural next step would be to extend this picture to $\sfZ_{U(M)}[y_1,\ldots,y_M]$ for $M\geq 3$ and to interpret it as the quantization of a Lagrangian variety in $(\mathbb{C}^*)^{2M}$. At present, however, we do not know how to carry out this construction systematically, even though for the models studied here we can derive both the explicit form of $\sfZ_{U(M)}[y_1,\ldots,y_M]$ and the corresponding operator $\hat{\sfA}_{U(M)}$. The technical complexity grows rapidly with $M$, making it increasingly difficult to infer directly the relevant Lagrangian variety and a complete set of quantum operators defining its quantization. A more conceptual understanding of the relation between the matrix-model difference equations, the associated Lagrangian geometry, and the corresponding structures in open topological strings therefore seems necessary.

The situation becomes considerably less clear once one moves beyond these three examples to more general $q$-deformed matrix models. In general, it is not known whether superintegrability persists, and the $q$-Virasoro constraints need not provide a well-defined recursive prescription for determining all correlators. Consequently, for a generic $q$-deformed matrix model we do not know how to construct systematically an operator $\hat{\sfA}_{U(1)}$ annihilating $\sfZ_{U(1)}[y]$. Nevertheless, on general grounds one may expect such an operator to exist. Even when a plausible candidate can be guessed, proving directly that it annihilates the corresponding matrix integral can still be highly nontrivial.

Finally, for generic $q$-deformed matrix models the choice of integration contour becomes an essential part of the problem, and one may expect some properties of the resulting difference equations to depend on this choice. This is closely analogous to the situation for non-Gaussian Hermitian matrix models, where the Virasoro constraints alone do not determine a unique solution.

\subsubsection*{Acknowledgements}

 We are deeply grateful to Pietro Longhi for his patient explanations and for numerous helpful discussions.
The research of V.\ M.\  is supported by the Priority 2030 Academic Leadership Initiative, contributing to the educational work of "Universities for a New Generation of Leaders", a project within the framework of the federal Youth and Children program. The research of M.\ Z.\ 
   is  supported by the Swedish Research Council excellence center grant ``Geometry and Physics'' 2022-06593. M.\ Z.\ is grateful to the 
   Simons Foundation for the financial support through the grant "Simons visiting program at GGI" and to GGI for the hospitality.

\appendix

\section{Operators at finite number of variables}\label{sec:AppendixDIMtoDAHA}
For a finite number $M$ of $\by$ variables we have
\begin{equation}
    x^\pm_k 
 = q^{\mp M}(\delta_{k,0}-(1-q^{\pm1})u^\pm_k)~,
\end{equation}
where 
\be
 u^\pm_{-k}:=\sum_{i=1}^M \y_i^k \prod_{j\neq i}\frac{q^{\pm1}\y_i-\y_j}{\y_i-\y_j}
 \T^{\pm1}_{i}~,
\ee
and 
\begin{equation}
p_{k}=\sum\limits_{i=1}^{M}y_i^k\,.
\end{equation}

\section{Proof of relations for partition functions.}\label{sec:AppendixSplittingProof}
First, we prove the contiguous relation for the rank one partition function and explain how to manipulate the expressions for the action of the $A^{\CS}_1$ operator on the rank two partition function.

The relation 
\begin{equation}\label{eq:contiguousappendix}
    \sfZ^{\CS}_{U(1)}[q^{-1 }y_i,a q^{-1}]- \sfZ^{\CS}_{U(1)}[y_i,a q^{-1}]= \dfrac{q-a}{q^2} y_i\sfZ^{\CS}_{U(1)}[y_i,a ] 
\end{equation} 
can be proven by considering the exact expression \eqref{CS-M2-PF} for the partition function. It is more instructive, however, to only use the quantum curve equations. The idea is to relate the operator $\hat{\sfA}_{U(1)}^{\CS}[\y,\T]$ to itself after shifting $a$. We have the following operator identity
\begin{equation}
    \hat{\sfA}_{U(1)}^{\CS}[\y,\T, a]\cdot(\T^{-1}-1)=(q\T^{-1}-1)\cdot   \hat{\sfA}_{U(1)}^{\CS}[\y,\T, aq^{-1}]~.
\end{equation}
Applying both sides to $\sfZ^{\CS}_{U(1)}[y_i, a q^{-1}]$, we find:
\begin{equation}
        \hat{\sfA}_{U(1)}^{\CS}[\y,\T, a] \cdot \left( \dfrac{ \sfZ^{\CS}_{U(1)}[q^{-1 }y,a q^{-1}]- \sfZ^{\CS}_{U(1)}[y,a q^{-1}]}{y}\right)=0~.
\end{equation}
Therefore the combination in parentheses is annihilated by the quantum curve. The solution to the quantum curve equation is unique up to a constant, hence this combination of shifted partition functions should be proportional to the partition function itself:
\begin{equation}
   \dfrac{ \sfZ^{\CS}_{U(1)}[q^{-1 }y,a q^{-1}]- \sfZ^{\CS}_{U(1)}[y,a q^{-1}]}{y}= {\rm const}  \times \sfZ^{\CS}_{U(1)}[y,a]
\end{equation}
Fixing the constant we obtain \eqref{eq:contiguousappendix}. 
We can rewrite it as the relation:
\begin{equation}\label{eq:contiguousappendix2}
   \sfZ^{\CS}_{U(1)}[y,aq^{-1}]- \dfrac{a-q}{q^2}y\sfZ^{\CS}_{U(1)}[y,a]= \sfZ^{\CS}_{U(1)}[yq^{-1},aq^{-1}]
\end{equation}
Collect the original partition function and the one with shifted $a$ into a vector:
\begin{equation}
    {V}(y)=\begin{pmatrix}
        y\sfZ^{\CS}_{U(1)}[y,a] \\
        \sfZ^{\CS}_{U(1)}[yq^{-1},aq^{-1}]
    \end{pmatrix}
\end{equation}
Evaluating the contiguous relations at $y$ and $qy$ we get that
\begin{equation}\label{eq:Vshift}
      {V}(qy)=M_{+}(y)  {V}(y)~, \quad  {V}(q^{-1}y)=M_{-}(y)  {V}(y) ~,
\end{equation}
for explicit $2\times 2$ matrices $M_{+}(y)$ and $M_{-}(y)$. The rank two partition function can be rewritten in terms of $V(y)$:
\begin{equation}
    \sfZ^{\CS}_{U(2)}[y_1,y_2]= \dfrac{V(y_1)^{T}J V(y_2)}{y_1-y_2}
\end{equation}
with $J=\begin{pmatrix}
    0 & 1\\-1 & 0
\end{pmatrix}$. Finally, after applying $\hat{\sfA}^{\CS}_{U(1)}[y_1]$ and using  \eqref{eq:Vshift} we get:
\begin{equation}
   \begin{split}
       &\hat{\sfA}^{\CS}_{U(1)}[y_1]\cdot \dfrac{V(y_1)^{T}J V(y_2)}{y_1-y_2}= \\
       &= V(y_1)^{T}\left( -\dfrac{1+q^{-1}y_1}{y_1-y_2}J+\dfrac{M_-(y_1)^TJ}{q^{-1}y_1-y_2}+\dfrac{ay_1}{q}\dfrac{M_+(y_1)^TJ}{qy_1-y_2}\right) V(y_2)
   \end{split}
\end{equation}
Computing the same for $  \hat{\sfA}^{\CS}_{U(1)}[y_2]$ we find that the two expressions are equal; hence the equation $\hat{A}_1^{\CS} \sfZ^{\CS}_{U(2)}[y_1,y_2]=0$.
\\

Next we give the proof of the splitting property \eqref{eq:splitCS} for the $\hat{\mathcal{O}}^{\CS}_2$ operator. Write the $U(2)$ partition function as a sum over two-row partitions:
\begin{equation}\label{eq:Z2CSapp}
    \sfZ^{\CS}_{U(2)}[y_1,y_2]= \sum_{n \geq m}c_{[n,m]}(a) s_{[n,m]}(y_1,y_2)~,
\end{equation}
where we highlighted the $a$-dependence of the coefficient. The two-row Schur functions of two variables are given by:
\begin{equation}
    s_{[n,m]}(y_1,y_2)=\dfrac{y_1^{n+1}y_2^m-y_1^{m} y_2^{n+1}}{y_1-y_2} = \sum_{l=0}^{n-m} y_1^{n-l}y_2^{m+l}~.
\end{equation}
The coefficients are explicitly known, by using the $q$-deformed hook formula to express the $s_{\lambda}\left(p_k=\dfrac{1}{1-q^k}\right)$ factor:
\begin{equation}\label{eq:tworowSI}
    c_{[n,m]}(a,q)=\dfrac{q^m\prod\limits_{j=1}^n q^{j-1}(1-a q^{j-1})\prod\limits_{j=1}^n \dfrac{q^{j-2}(1-a q^{j-2})}{(1-q^{n-j    +1})(1-q^{m-j+1})} }{\prod\limits_{j=m+1}^{n}(1-q^{n-j+1})}
\end{equation}
After acting on the series \eqref{eq:Z2CSapp} and collecting the coefficients in front of monomials $y_1^n y_2^m$ we get:
\begin{equation}
    \hat{\mathcal{O}}^{\CS}_2\sfZ^{\CS}_{U(2)}[y_1,y_2] = \sum_{n,m} \alpha_{n,m} y_1^n y_2^m~,
\end{equation}
where
\begin{equation}
    \alpha_{n,m}=a q^{n+m-2} \sum_{k=0}^{m-1} c_{[n+k,m-k-1]}(a,q)-\sum_{k=0}^m c_{[n+k,m-k]}(a,q)~.
\end{equation}
We would like to compare this expression to the expansion of the RHS of \eqref{eq:splitCS}
\begin{equation}
    -  \sfZ_{U(1)}^{\rm CS}[y_1]\sfZ_{U(1)}^{\rm CS}[y_2\, |\, a q^{-1}]= \sum_{n,m} \left(-c_{[n]}(a,q)c_{[m]}(a q^{-1},q) \right) y_1^n y_2^m~.
\end{equation}
At this point it is simpler to consider the following combination:
\begin{equation}
    \alpha_{n,m}-\alpha_{n-1,m+1}=c_{[n-1,m+1]}-a q^{n+m-2}c_{[n-1,m]}~.
\end{equation}
Hence the equality we should check turns into simply
\begin{equation}
    c_{[n,m+1]}(a)-a q^{n+m-1}c_{[n,m]}(a) = c_{[n+1]}(a)c_{[m]}(a q^{-1})-c_{[n]}(a)c_{[m+1]}(aq^{-1})~.
\end{equation}
This can be shown by using the explicit expression \eqref{eq:tworowSI}, canceling the common factors and comparing the resulting expressions.

\section{Derivation of mirror curves}\label{app:mirror}

In this appendix we explain how the classical curves for different matrix models are related to 
 the mirror curves for the toric Calabi-Yau manifolds with Hori-Vafa mirror symmetry \cite{Hori:2000kt}. 
  The case of the Chern-Simons curve $\Sigma^{\rm CS}_a$ defined in \eqref{CS-classical-curve} is well-known 
   and it appears as the part of the mirror for the resolved conifold which is a $\mathbb{C}^4//U(1)$ K\"ahler quotient 
    with the charge matrix $(1,1,-1,-1)$. Since this is a very well-known example, we will not discuss it here and 
     will concentrate on two other matrix model's curves.

\subsection{$q$-Laguerre curve}\label{app:mirror-qL}

Here we provide details on the mirror symmetry interpretation of the curve \eqref{qL-general-qcurve}.
Consider its classical counterpart in $(X_1, X_2) \in (\mathbb{C}^*)^2$
  \begin{equation}
      1+ X_1 + X_2 + a (r+1) X_1 X_2+ a^2 r  X_1 X_2^2=0~.
  \end{equation}
 This is a mirror curve for a specific toric CY manifold $\mathbb{C}^5//U(1)^2$ which we describe below. 
 Consider $\mathbb{C}^5$ and two $U(1)$-actions defined by the following charge matrix
   \begin{equation}
\begin{pNiceArray}{ccccc}[first-row]
 z_1 & z_2 & z_3 & z_4 & z_5 \\
 1   & 1   & -1  & -1  & 0   \\
 1   & -1  & -1  & 0   & 1
\end{pNiceArray}~,
\end{equation}
  where $t_1$ and $t_2$ would be the K\"ahler parameters for the first and second lines of charges, respectively. 
 This geometry corresponds to the resolution of the suspended pinch point. To see this we can introduce
  the coordinates 
  \begin{equation}
    w_1 = z_1 z_3~,~~~w_2=z_2 z_4 z_5~,~~~w_3= z_2 z_3 z_5^2~,~~~w_4=z_1 z_2 z_4^2  ~,
  \end{equation}
  such that in $\mathbb{C}^4$ we have
   \begin{equation}
       w_1 w_2^2 = w_3 w_4~. 
   \end{equation}
 Now let us derive the Hori-Vafa mirror curve for this geometry. 
  Assuming $Q_i= e^{t_i}$
   the Hori-Vafa mirror curve is defined by the equations
   \begin{align}
       & y_1 + y_2 + y_3 + y_4 + y_5 =0~,~~~y_i \in \mathbb{C}^*~,\nonumber \\
       & y_1 y_2 y_3^{-1} y_4^{-1} = Q_1~,\\
       & y_1 y_2^{-1} y_3^{-1} y_5 = Q_2~.\nonumber
   \end{align}
    If we switch to homogeneous coordinates 
    \begin{equation}
        X_2= \frac{y_3}{y_1}~,~~~~X_1 = \frac{y_4}{y_1}~,
    \end{equation}
    and resolve the  constraints we get
    \begin{equation}
      1 + X_1 + X_2 + Q_1 X_1 X_2 + Q_1 Q_2 X_1 X_2^2=0~.  
    \end{equation}
  For the curve $ \Sigma^{\rm qL}_{a, r}$ the parameters are related as 
     \begin{equation}
        a(r+1)= Q_1~,~~~~a^2 r = Q_1 Q_2~.
     \end{equation}
     The choice $r=1$ corresponds to $4Q_2=Q_1$. 

\subsection{$q$-Gaussian curve}\label{app:mirror-qG}

 Here we provide the mirror symmetry interpretation for the quantum curve \eqref{qcurve-qG-general} for the $q$-Gaussian model. 
  The classical curve for \eqref{qcurve-qG-general} corresponds to the following expression for $(X_1, X_2) \in (\mathbb{C}^*)^2$
\begin{align}\label{qG-clas-curve-gen}
      1 + X_1 + X_2 + a X_1 X_2
     + R X_1^2 + 2 aR X_1^2 X_2 +  a^2 R X_1^2 X_2^2=0 ~.
   \end{align}
  Consider the toric Calabi-Yau manifold $\mathbb{C}^7//U(1)^4$
  with the following matrix of charges
  \begin{equation}
      \left ( \begin{array}{ccccccc}
         1  & -1 & -1 & 1 & 0 & 0 & 0 \\
          1 & -2 & 0 & 0 & 1 & 0 & 0 \\
          0 & 1 & 0 & -1 & -1 & 1 & 0\\
          0 & 0 & 0 & 0 & 1 &-2  & 1
      \end{array} \right )~,
  \end{equation}
  where every line has the corresponding K\"ahler parameter $t_i$ and we define $Q_i= e^{t_i}$.
  Thus the Hori-Vafa mirror curve is defined by the equations
   \begin{align}
       & y_1 + y_2 + y_3 + y_4 + y_5 + y_6 + y_7 =0~,~~~y_i \in \mathbb{C}^*~,\nonumber \\
       & y_1 y_2^{-1} y_3^{-1} y_4 = Q_1~, \nonumber\\
       & y_1 y_2^{-2}  y_5 = Q_2~, \\
       & y_2 y_4^{-1} y_5^{-1} y_6 = Q_3~,\nonumber \\
       & y_5 y_6^{-2} y_7 = Q_4~.\nonumber
   \end{align}
    We define the coordinates 
    \begin{equation}
        X_1 = \frac{y_2}{y_1}~,~~~~X_2 = \frac{y_3}{y_1}~.
    \end{equation}
  Then if we resolve these constraints we get the following curve
  \begin{equation}
      1 + X_1 + X_2 + Q_1 X_1 X_2 + Q_2 X_1^2 + Q_1 Q_2 Q_3 X_1^2 X_2 + Q_1^2 Q_3^2 Q_2 Q_4 X_1^2 X_2^2=0
  \end{equation}
 The classical curve \eqref{qG-clas-curve-gen} for the $q$-Gaussian model corresponds to the following choice of parameters
 \begin{equation}
     Q_1 = a~,~~~~Q_2 =R~,~~~~Q_3 = 2~,~~~~Q_4 = \frac{1}{4}~.
 \end{equation}

\section{Birational anti-symplectomorphisms}\label{app:birational}

In this appendix we want to explain how to get birational anti-symplectomorphisms which preserve the curve and also derive the map defined by the expressions \eqref{eq-qG-invol-extra1} and \eqref{eq-qG-invol-extra2}. We consider the concrete case of the $q$-Gaussian curve, but similar considerations can be carried out for other curves. 

Remember that the classical curve for the $q$-Gaussian model is defined by the following expression
  \begin{equation}
      \mathcal{A}^{\rm qG} (y,T) = 2 ay^2  -1 + (1-y^2)T^{-1} - a^2 y^2 T~.
  \end{equation}
  Let us introduce new coordinates
   \begin{equation}\label{qG-new-coord}
       z = y(1-aT)~,~~~~s= \mathcal{A}^{qG}+1 =\frac{1- y^2(1-aT)^2}{T}~,
   \end{equation}
  or its inverse
  \begin{equation}
      T= \frac{1-z^2}{s}~,~~~~y = \frac{sz}{s-a+ az^2}~.
  \end{equation}
   These are birational transformations and they do not map the entire $(\mathbb{C}^*)^2$ to $(\mathbb{C}^*)^2$,
    since some divisors should be removed. One can check the following property
    \begin{equation}
    \omega=    \frac{dz}{z} \wedge \frac{ds}{s} = \frac{dT}{T}\wedge \frac{dy}{y}~,
    \end{equation}
    thus this is a birational symplectomorphism of $(\mathbb{C}^*)^2$. If we work in $(s,z)$ coordinates we have the following infinite family of 
      anti-symplectomorphisms
      \begin{equation}
          \iota (s, z) = \Big (s, \frac{\alpha(s)}{z} \Big )
      \end{equation}
 where $\alpha(s)$ is some rational function.  Such maps are anti-symplectomorphisms $\iota^* (\omega) = - \omega$ and involutions: $\iota^2=1$.  
 In these adapted coordinates the curve corresponds to the equation $s=1$ and thus the map obviously preserves the curve. We can rewrite the map $\iota$ in $(y,T)$ coordinates and thus obtain a birational 
 anti-symplectomorphism preserving $\mathcal{A}^{\rm qG}$.  
 
 Let us choose $\alpha$ to be a constant $\alpha = \frac{a-1}{a}$.
 Then using \eqref{qG-new-coord} the involution $\iota$ in terms of original coordinates 
   \begin{equation}
       \iota (y, T) = (y', T')
   \end{equation}
  has the following form
  \begin{align}
 & T' = T \frac{\rho-\alpha^2}{\rho (1-\rho)}~,\\
   &    y'=\frac{\alpha y(1-aT)(1-\rho)}{
\rho(1-\rho)-aT(\rho-\alpha^2)} ~,
      \end{align}
   where we have used the shorthand notation $\rho=y^2 (1-aT)^2=z^2$. 
  We can show the following properties
  \begin{equation}
      \iota^* \Big ( \frac{dT}{T} \wedge \frac{dy}{y} \Big ) =\frac{dT'}{T'} \wedge \frac{dy'}{y'}= -  \frac{dT}{T} \wedge \frac{dy}{y}
  \end{equation}
 and 
 \begin{equation}
   \mathcal{A}^{\rm qG} (y,T) =  \mathcal{A}^{\rm qG} (y',T')~.    
 \end{equation}

\section{Classical limit of operators}

It is also possible to extract differential equations for classical matrix models by taking the direct $\hbar \rightarrow 0$ limit at fixed $N$. Let us denote the respective partition functions of the Gaussian and Laguerre models as $\sfZ^{\rm G}_{U(M)}[y_1,\ldots ,y_M]$ and $\sfZ^{\rm L}_{U(M)}[y_1,\ldots ,y_M]$.
 Just as in the $q$-deformed case they are averages of multi-point functions of inverse characteristic polynomials in the respective  ensembles.
One has to go to the first order in the expansion
\begin{equation}
    \hat{T}=\exp\left( {\hbar y \frac{\partial}{\partial y}} \right)= 1+ \hbar \hat{D} \,,\quad \hat{D}=y \frac{\partial}{\partial y}~.
\end{equation}
To obtain the expansions directly, we also notice that:
\begin{equation}
    1-q^{n}a^{m}\hat{T}^{k} \rightarrow -\hbar\left( n+mN+k  \hat{D}\right)
\end{equation}
The Chern-Simons model doesn't have an interesting classical limit, hence we consider the two other examples
 \begin{equation}
     q^{-1}\T \hat{\sfA}^{\rm qL}_{U(1)} = (1-\T)-\y(1-a \T)(1- a\T)
 \end{equation}
and
 \begin{equation}
     q^{-2}\T \hat{\sfA}^{\rm qG}_{U(1)} = (1-\T)-\y^2(1-a \T)(1-q a\T)~.
 \end{equation}
Note that, in order to make the limit well defined, according to formulas \eqref{eq:scaling-weight} and \eqref{eq:SIlimit} we have to rescale $y_i \rightarrow \dfrac{y_i}{\hbar}$ for the $q$-Laguerre model and $y_i \rightarrow \dfrac{y_i}{\hbar^{1/2}}$ for the $q$-Gaussian
\begin{align}
 & \sfZ^{\rm L}_{U(M)}[y_1, ..., y_M, N]=  \lim\limits_{\hbar \rightarrow 0} \sfZ^{\rm qL}_{U(M)}[y_1\hbar^{-1},..., y_M\hbar^{-1}, q, a]~,\\
 & \sfZ^{\rm G}_{U(M)}[y_1,...,y_M, N]=  \lim\limits_{\hbar \rightarrow 0} \sfZ^{\rm qG}_{U(M)}[y_1\hbar^{-1/2},...,y_M\hbar^{-1/2}, q, a]~.
\end{align}

This scaling is consistent with the structure of the recursion operator and after implementing it all terms appear to be of order $\hbar$, hence:
\begin{equation}
     q^{-2}\T \hat{\sfA}^{\rm qG}_{U(1)} \rightarrow \hat{D}-y^2 \left( \hat{D}+N\right)\left(\hat{D}+N+1 \right)\,
\end{equation}
and
\begin{equation}
     q^{-1}\T \hat{\sfA}^{\rm qL}_{U(1)} \rightarrow \hat{D}-y \left( \hat{D}+N\right)\left(\hat{D}+N \right)\,.
\end{equation}

Using these equations we can recover the standard large $N$ expansion. After rescaling $y$ by a factor of $1/\sqrt{N}$ we get
\begin{equation}\label{eq:LargeNGauss}
   \left( \hat{D}-\dfrac{y^2}{N} \left( \hat{D}+N\right)\left(\hat{D}+N+1 \right)  \right)\sfZ^{\rm G}_{U(1)}[y, N]=0~.
\end{equation}
At the leading order the partition function behaves as
\begin{equation}
    \sfZ^{\rm G}_{U(1)}[y] \sim \exp\left(N \sfF_0^{\rm G}[y] \right)~.
\end{equation}
Substituting the asymptotics into equation \eqref{eq:LargeNGauss}  we get an algebraic equation for the derivative of the free energy $W(y) = y \dfrac{dF_0^{\rm G}(y)}{dy}$:
\begin{equation}
    -y^2 W(y)^2+W(y)(1-2 y^2)-y^2= 0 
\end{equation}
This equation is nothing but the leading order loop equation and is solved by
\begin{equation}
    W(y)=\frac{1-2 y^2-\sqrt{1-4 y^2}}{2 y^2}=c(y^2)-1~,
\end{equation}
where $c(y^2)$ is  the Catalan generating function reproducing the standard result. One can go through a similar analysis for the Laguerre model.

Exactly the same procedure can be applied to the rank two recursion operator. The result will be nothing but the cut-and-join equation of the Gaussian model expressed in the $y$ variables. 
More interesting is the limit of the quantum variety equations. Since we have already covered $\hat{\sfA}^{\rm }_{U(1)}$ we need to study what happens to the splitting operators. We will only treat the $q$-Laguerre case for simplicity. Recall that we have
\begin{equation}
     \hat{\mathcal{O}}^{\qL}_2= \y_2 (1-a^2 q^{-1}\T_1 \T_2)-1
\end{equation}
After proper rescaling we have
\begin{equation}
   \hat{\mathcal{O}}^{\rm L}_2=   y_2 \left( \hat{D}_1+\hat{D}_2+2N-1   \right)-1 \,
   ,\quad \hat{D}_i= y_i\dfrac{\partial}{\partial y_i}~,
\end{equation}
 and we define $\hat{\mathcal{O}}^{\rm L}_1$ by exchanging the labels $1$ and $2$. 
  Thus from \eqref{eq-qL-full-pre}-\eqref{eq-qL-full} we obtain three PDEs for $\sfZ^{\rm L}_{U(2)}[y_1, y_2, N]$
  \begin{align}
& \Big (\hat{D}_1-y_1 \left( \hat{D}_1+N\right)\left(\hat{D}_1+N \right) \nonumber\\
&~~~~~~~~~~~~~~~
 -\hat{D}_2-y_2 \left( \hat{D}_2+N\right)\left(\hat{D}_2+N \right) \Big )\sfZ^{\rm L}_{U(2)}[y_1,y_2, N]=0~,\\
  &    \left( \hat{D}_1-y_1 \left( \hat{D}_1+N\right)\left(\hat{D}_1+N \right) \right) \hat{\mathcal{O}}^{\rm L}_2 \sfZ^{\rm L}_{U(2)}[y_1,y_2, N]=0~,\\
 &   \left( \hat{D}_2-y_2 \left( \hat{D}_2+N\right)\left(\hat{D}_2+N \right) \right) \hat{\mathcal{O}}^{\rm L}_1 \sfZ^{\rm L}_{U(2)}[y_1,y_2,N]=0~.
  \end{align}
  One can repeat the same analysis for the Gaussian model and write 
   similar formulas for the Gaussian model. 
  The determinant formula of \eqref{qL-M2-PF} in the limit becomes:
\begin{equation}
    \sfZ^{\rm L}_{U(2)}[y_1,y_2,N]=\dfrac{1}{y_1-y_2}\det \begin{pmatrix}
        y_1\sfZ^{\rm L}_{U(1)}[y_1,N] &  y_2\sfZ_{U(1)}^{\rm L}[y_2,N]\\
          \sfZ^{\rm L}_{U(1)}[y_1,N-1]&   \sfZ_{U(1)}^{\rm L}[y_2,N-1]
    \end{pmatrix} 
\end{equation}
 We can write a similar formula for the Gaussian model.

\bibliographystyle{utphys}
\bibliography{references,references2}
\end{document}